\documentclass{article}

\usepackage[eandd,final]{neurips_2026}

\usepackage[utf8]{inputenc}
\usepackage[T1]{fontenc}
\usepackage{float}
\usepackage{placeins}
\usepackage{hyperref}
\hypersetup{hidelinks}
\usepackage{url}
\usepackage{booktabs}
\usepackage{amsmath}
\usepackage{amssymb}
\usepackage{microtype}
\DisableLigatures{encoding = *, family = tt*}
\usepackage{xcolor}
\usepackage{graphicx}
\usepackage{enumitem}
\usepackage{needspace}
\usepackage{multirow}
\usepackage{algorithm}
\usepackage{algpseudocode}

\newcommand{\bnf}{\textsc{Build-and-Find}}
\newcommand{\acc}{\mathrm{Acc}}

\title{\bnf{}: An Effort-Aware Protocol for Evaluating Agent-Managed Codebases}

\author{
  Jhen-Ke Lin \\
  National Yang Ming Chiao Tung University \\
  \texttt{jacob.cs14@nycu.edu.tw}
}

\begin{document}

\maketitle

\begin{abstract}
Most coding-agent benchmarks ask whether generated code behaves correctly. That
remains essential, but repository-level engineering is increasingly
agent-managed: one agent writes a repository, and later agents inspect, audit,
or extend it as working context. In that setting, a generated repository is not
only an answer to a task but also a communication artifact for future work. Even
when strong agents nearly satisfy the visible behavioral objective, repositories
can differ in how clearly they expose the intended behavior and design choices
behind that behavior. We introduce \bnf{} to evaluate whether downstream agents
can efficiently recover intended choices from generated repositories. For each
task, a builder creates a codebase from a hidden specification; a finder receives
only the codebase and a specification-traced question bank.
The protocol separates behavioral correctness from artifact-side recovery and
reports recovery accuracy, repeatability, implementation coverage, and
inspection effort. Evaluation proceeds in two stages: the first uses recovery
accuracy and stability as gates to establish that finders can reliably recover
the intended choices; the second compares inspection effort by the same finder
across artifacts that pass these gates and allow the same intent to be recovered.
Lower effort suggests that the artifact makes that intent easier to locate.
We find that several frontier models approach saturation in intent recovery on
our task pack, yet second-stage inspection effort reveals differences among
their generated artifacts, providing a dimension beyond accuracy for comparing
artifact legibility. We also audit artifacts to distinguish unimplemented
specification claims from finder failures and check whether correct answers
are supported by artifact evidence. We release the harness and benchmark
data for reproducibility and extension.
\end{abstract}

\section{Introduction}

Passing tests establishes behavioral evidence without guaranteeing that code is
easy to understand. Empirical work on ChatGPT-generated programs already treats
correctness and code quality as separable concerns: beyond checking behavioral
outcomes, Liu et al. analyze style and maintainability issues in generated Java
and Python programs \citep{liu2024refining}. As coding agents move from
function synthesis to repository-level work, generated codebases increasingly
become intermediate artifacts: one agent writes them, another inspects them, and
later agents audit, modify, or extend them. A repository that hides design
intent can impose extra inspection burden even when visible tests pass; one that
exposes intent can make the same design easier to recover.

This mirrors a familiar software-engineering principle: maintainable code
combines correctness with structure that helps future readers understand why it
is written that way. Readability and structure are established software-quality
concerns \citep{buse2010readability,borstler2023codequality}, and Parnas's
modularization criteria frame decomposition as a tool for system
comprehensibility \citep{parnas1972criteria}. In agent-managed repositories,
the future reader may be another agent. The corresponding question is whether
the repository exposes intended behavior and design choices clearly enough for
later agents to inspect, audit, and use it as context.

We separate three layers of agent-written software evaluation: behavioral
correctness, artifact-side intent recoverability, and downstream modification
success. Existing coding benchmarks primarily target behavioral correctness,
including unit-test passing in HumanEval and MBPP
\citep{chen2021codex,austin2021program}, harder programming tasks in APPS and
DS-1000 \citep{hendrycks2021apps,lai2023ds1000}, and issue-resolution patches
in SWE-bench \citep{jimenez2024swebench}. \bnf{} targets the middle layer:
before asking whether a later agent can safely modify a repository, we ask
whether it can recover the intended behaviors and design choices from the
repository itself.

The target construct is artifact-evidenced recovery. A correct answer should be
distinguishable from alternatives using evidence in generated source files,
tests, configuration, or documentation. This gives artifact-side evaluation a
sharp boundary. Functional correctness asks whether the repository behaves
correctly; downstream modification success asks whether a later agent can safely
change it; artifact-evidenced recovery asks whether intended decisions are
visible in the artifact before any modification is attempted.

In high-recovery regimes, accuracy alone is a coarse measure of artifact
legibility. If two generated repositories both allow downstream agents to
recover the same intended decisions, the amount of inspection required for that
recovery becomes informative. An artifact that exposes intent through clear
structure, tests, configuration, or documentation should require less finder
inspection than one that buries the same intent. \bnf{} therefore treats
inspection effort as an agent-facing proxy for artifact legibility, interpreted
only after recovery and stability gates.

The central idea is to treat generated code as external memory and as a
communication medium. A hidden specification defines intended behaviors and
design choices; a builder encodes them in a codebase; and a finder, given only
the codebase and a question bank, tries to recover them. Accurate recovery with
low inspection effort, after stability checks, indicates that the repository
made the intended choices easier for downstream agents to find. The
builder-finder split separates implementation ability from artifact recovery
while exposing interactions between artifacts and inspection policies.

This paper makes four contributions.
\begin{enumerate}[leftmargin=1.25em,itemsep=0.18em,topsep=0.18em]
  \item \textbf{Evaluation target.} We define artifact-evidenced recovery of
  intended repository behavior and design choices as a distinct layer between
  behavioral correctness and downstream modification success.
  \item \textbf{Builder--finder protocol.} Builders receive hidden
  specifications and produce repositories; finders are given only those
  repositories plus specification-traced question banks.
  \item \textbf{Two-stage evaluation.} We first assess recovery accuracy and
  stability, then compare inspection effort as a proxy for artifact legibility,
  reporting implementation coverage alongside both stages.
  \item \textbf{Calibration and release.} We include question-only, spec-only,
  compile-failed, low-prior, evidence-audit, and builder--finder affinity
  diagnostics, and release the harness, artifacts, records, canonical tables,
  reports, scripts, metadata, licenses, and evidence audits needed for audit.
\end{enumerate}

\paragraph{Empirical scope.}
\bnf{} treats controls as part of the evaluation contract: they separate
ordinary engineering priors, specification access, and artifact-conditioned
recovery. Section~\ref{sec:prior_stress} reports the calibration numbers; the
empirical claims in this release are panel-local and prior-conditioned.

\section{Related work}

\paragraph{Functional code generation.}
HumanEval \citep{chen2021codex} and MBPP \citep{austin2021program} established
compact unit-test-based evaluations for code models. APPS
\citep{hendrycks2021apps}, DS-1000 \citep{lai2023ds1000}, and LiveCodeBench
\citep{jain2025livecodebench} broaden the task distribution and contamination
controls. These benchmarks answer whether generated code satisfies behavioral
tests. Our protocol asks a different question: after a repository is generated,
does it expose the intended decisions clearly enough for a downstream agent to
recover them?

\paragraph{Code understanding and code QA.}
Code-understanding and code-QA benchmarks evaluate adjacent capabilities.
CodeXGLUE \citep{lu2021codexglue} combines program-understanding and generation
tasks, InfiBench \citep{li2024infibench} evaluates code-related QA, and
repository-level QA benchmarks such as SpyderCodeQA, CodeRepoQA, and SWE-QA
test semantic, dependency, intention, and multi-hop repository understanding
\citep{strich2024repoqa,hu2025coderepoqa,peng2025sweqa}. These tasks usually
start from an existing repository. \bnf{} first asks a builder to create a
repository from a hidden specification, then asks what information that
generated artifact carries for a finder that is not given the specification.

\paragraph{Repository-level software engineering.}
SWE-bench \citep{jimenez2024swebench} and SWE-bench Verified
\citep{openai2024swebenchverified} evaluate model patches on real GitHub
issues. RepoBench \citep{liu2024repobench} targets repository-level retrieval
and completion. These settings capture realistic software context through patch
correctness or completion endpoints. Our endpoint is different: the produced
repository itself becomes the object that future agents must inspect.

\paragraph{Program comprehension and maintainability.}
Program-comprehension and maintainability work studies how readers ask
questions about code and reason about behavior
\citep{sillito2006questions,ko2008debugging,ko2010whyline,buse2010readability,oliveira2020readability}.
\bnf{} uses this perspective to define an intermediate evaluation target. Before
measuring whether a later agent can safely modify a generated repository, we
measure whether it can recover the decisions that the repository was meant to
express.

\paragraph{Interactive and long-horizon agent benchmarks.}
Interactive and long-horizon benchmarks such as $\tau$-bench, MLE-bench,
PaperBench, and Terminal-Bench evaluate agents that act through tools, APIs,
terminals, research workflows, and sandboxed environments
\citep{yao2025taubench,chan2025mlebench,starace2025paperbench,merrill2026terminalbench}.
These benchmarks share our focus on agentic, tool-mediated work. Their primary
outcome is task completion in an environment; our benchmark asks whether the
artifact produced by one agent makes intended decisions recoverable for another
agent, and how much inspection that recovery requires.

\section{Protocol design}
\label{sec:design}

\subsection{Builder-finder protocol}

Each task $t$ contains a hidden specification with intended behaviors and design
choices. It also contains a question bank whose gold answers are traced to
specific items in that specification. These traces make the recovery target
auditable, following the broader role of traceability in software-intensive
systems \citep{clelandhuang2014traceability}. The question-bank form is also
motivated by program-comprehension work showing that developers ask concrete
questions during evolution tasks \citep{sillito2006questions}, including why
and why-not questions about program behavior
\citep{ko2008debugging,ko2010whyline}.

A builder $b$ receives the task prompt and constructs a self-contained codebase
$A_{b,t,r}$ on trial $r$. Build validation records compile checks and structural
metadata. A finder $f$ is then given only the artifact and the question bank; the
hidden specification is withheld from its prompt and workspace. Functional tests ask whether the code
behaves as intended. \bnf{} asks whether the intended behavior and design
choices are recoverable from the artifact. In the reported study, each of the 12
builder configurations generates two repositories per task from the hidden
specification, and every compile-passing repository is inspected by all 12
finder configurations over three trials
(Algorithms~\ref{alg:bench}--\ref{alg:finder} in Appendix~\ref{app:procedure}).

The finder inspects files, makes tool calls, and writes a structured answer
file. Each question is a four-option multiple-choice item with deterministic
option shuffling per run. For example, one item asks how a web framework
composes its middleware stack; its gold option, a list of trait objects whose
registration order drives outer-to-inner call order, traces to a single
specification requirement, and a finder can answer it by locating the
registration and dispatch code (Appendix~\ref{app:procedure} lists the full
item). Exact-match grading avoids free-form answer parsing and keeps partial
runs auditable.

The intended construct is artifact-evidenced recovery: a correct answer should
be distinguishable from plausible alternatives by evidence in source files,
tests, configuration, or documentation. Exact-match grading keeps the formal
score deterministic. Specification traces, prior controls, compile-failed stress
diagnostics, and the post-hoc artifact-evidence audit in
Appendix~\ref{app:evidence-audit} test whether that deterministic score matches
the intended construct. The audit counts a correct answer as evidence-supported
only when the cited file, symbol, or snippet-level rationale distinguishes the
selected option from plausible alternatives.

We refer to these artifact-conditioned runs as the \emph{formal condition}; some
tables abbreviate them as formal runs or rows.

\subsection{Finders as measurement instruments}

Finders play two roles. In the benchmark, they are calibrated measurement
instruments. In the target deployment setting, they approximate downstream
agents that must use generated repositories as external memory. The measured
result therefore depends on both the artifact and the finder: a weak finder can
make good artifacts look opaque, while a very strong finder can mask artifact
differences. This mirrors a broader evaluation problem: assessing LLM-generated
code and text is difficult, and LLM-as-a-judge methods require calibration for
reliable use \citep{wang2025llmjudge}.

We therefore analyze finder validity before comparing artifacts. A useful finder
panel has high artifact-conditioned recovery, passes control checks, retains
performance on low-prior or implementation-signal questions, and spans more than
one model family. We also report builder--finder affinity: pair-specific
recovery or effort effects after accounting for the builder's average artifact
legibility and the finder's average inspection ability. Affinity is panel-local
and diagnostic. Appendix~\ref{app:finder_calibration} reports the per-finder
calibration.

\subsection{Prior and specification controls}
\label{sec:controls}

Software tasks contain legitimate engineering priors. A finder may infer that a
database should use write-ahead logging or that a web server should separate
routing from middleware without reading a particular artifact. The benchmark
therefore quantifies prior answerability with controls while preserving natural
requirements.

In the \emph{question-only} control, a finder is given only the question bank.
In the \emph{spec-only} control, the finder receives the full task YAML,
including the \texttt{intentions} block. The specification is thus withheld from artifact-conditioned finders and
deliberately provided in this control. These controls give artifact-conditioned
accuracy its prior and specification context. In this task pack, question-only
exact-match accuracy is high, so accuracy serves primarily as a recovery gate
and calibration signal; conditional inspection effort carries the main
artifact-legibility comparison.
The effort analysis then asks a narrower question: among successful recoveries
on compile-passing artifacts, how much inspection does the same finder spend to
recover the same intended decisions? Compile-failed artifacts are reported as
stress diagnostics.

\subsection{Recovery and inspection-effort metrics}
\label{sec:metrics}

Evaluation proceeds in two stages: we first assess whether intended choices can
be recovered accurately and repeatably, then compare inspection effort within
the same finder-task context. Among artifacts encoding the same intended design,
lower effort suggests that the design is easier for the finder to locate.
Recovery accuracy, repeatability, and coverage remain part of this interpretation.

Finder recovery metrics use an audited scoring set: the claims that the
implementation audit labels as implemented with the gold specification behavior.
Let $Q_t$ be the question set for task $t$. For
an artifact $A_{b,t,r}$, let $Q^+_{b,t,r}\subseteq Q_t$ be the subset of
questions whose artifact-question audit label is implemented-gold. Non-gold and
ambiguous pairs enter builder-side implementation coverage reports and are
excluded from finder recovery scoring.
Per-run answer accuracy is
\begin{equation}
  \acc(b,f,t,r) =
  \frac{1}{|Q^+_{b,t,r}|}\sum_{q \in Q^+_{b,t,r}}\mathbf{1}\{\hat{y}_{b,f,t,r,q}=y_{t,q}\}.
\end{equation}
Here $\hat{y}$ is the finder's answer and $y$ is the gold option. Accuracy is
reported on this audited scoring set and, with $Q_t$ in place of $Q^+_{b,t,r}$,
on all answer rows; only all-row accuracy is compared with the question-only
control. A run is
\emph{all-correct} when all questions in $Q^+_{b,t,r}$ are answered correctly.
The observed inspection effort for a run is kept unchanged: the finder spent
the full recorded inspection bytes, while all-correct status is computed over
the audited scoring set. Accuracy is a validity check and context
signal: it says whether recovery happened, while artifact legibility is
evaluated through conditional inspection effort.

The reported effort unit is \emph{novel agent inspection bytes}: new artifact
bytes retrieved or requested by the finder through retrieval and tool calls,
excluding system prompts, tool schemas, transport overhead, and replayed
context. Bytes provide a tokenizer-independent unit; tokenization is
model-specific, so the same retrieved source span can yield different token
counts for different finders.
Because raw bytes are finder-specific, builder artifacts are compared only
within the same finder-task context. Conditional $R_b$ is the geometric mean,
over contributing all-correct finder-task cells, of the builder's mean
all-correct effort divided by the minimum contributing builder effort in that
same finder-task cell. Missing all-correct cells are excluded from $R_b$ and
reported as coverage failures.

We also report a task-level recovery gate. For each builder--finder--task cell,
the gate uses simple mean audited answer recovery over repeated find runs. The
number of contributing cells is reported as a separate coverage quantity. This
keeps the recovery gate separate from inspection effort; it asks whether
recovered answers are reliable where the artifact exposes implemented-gold
claims. Appendix~\ref{app:metric_details} gives the full byte accounting and
metric equations.

The three reported views answer different methodological questions. Exact-match
accuracy asks whether recovery occurred at all. Conditional $R_b$ asks, among
successful recoveries, how much inspection effort the same finder needed for
each artifact. The task-level recovery gate asks whether the cell is recoverable
enough for effort to be interpreted. Reporting all three avoids treating a
low-effort but unreliable artifact as equivalent to a consistently recoverable
one.

\section{Empirical validation}
\label{sec:empirical}

We report controls before effort so artifact legibility is interpreted only
after prior, coverage, and recovery gates. The empirical section therefore
starts with prior and stress calibration, then reports repeatability,
implementation-aware recovery, conditional effort, rank stability, affinity, and
low-prior evidence audits.

\subsection{Study setup}
\label{sec:setup}

The reported study uses two executable repository task families,
\texttt{scratch\_minidb} and \texttt{scratch\_nanoweb}, with 15 questions per
task. We evaluate Claude Opus 4.7 and Sonnet 4.6, GPT-5.5 and
GPT-5.4-mini, MiMo-v2.5-pro, and MiMo-v2.5, each at high and low reasoning
effort. Each builder-task cell has two build trials; each status-ok artifact
whose release path exists is inspected by each finder for three trials.

The artifact-present release contains 48 builds, 1728 artifact-conditioned find
records, and 25920 raw question rows. The compile-pass primary panel contains
41 builds, 1476 find records, and 22140 raw question rows. Seven
artifact-present builds failed the compile probe and are retained only as
stress-test diagnostics. Eleven of the 72 question-only control runs read files
outside their workspaces; excluding them changes question-only accuracy from
94.5\% to 97.2\% (Appendix~\ref{app:details}).

After the compile and artifact-presence checks, we audited each
artifact-question pair and labeled whether the generated artifact implements the
gold specification claim being scored. The audit uses a two-stage process. First,
an automated conformance-triage script compares stable answers from selected
base finders, \texttt{codex\_gpt5\_5} and
\texttt{claude\_code\_opus\_4\_7}, over repeated artifact-conditioned runs; both
selected base finders are also members of the reported finder panel.
An implemented-gold consensus prelabel requires both selected high-effort finders
to choose gold in at least two of three runs each. Second, rows in the manual-review queue are
inspected and merged as manual-review labels. The final release records this
provenance in \texttt{label\_source}: 681 of 720 artifact-question rows are
\texttt{consensus\_prelabel}, 24 are \texttt{manual\_review}, and 15 are
\texttt{rule\_empty\_artifact} (Appendix~\ref{app:artifact-layout-sensitivity}).

The audit labels pairs as implemented-gold, implemented-non-gold, or
absent/ambiguous. Finder recovery metrics are computed on implemented-gold
pairs; non-gold and ambiguous pairs are reported separately as builder-side
implementation coverage. The audit constructs the scoring set to avoid
penalizing finders for gold claims that a builder artifact did not implement.
In the compile-pass primary panel, this gives 592 audited scoring pairs
and 21312 scored finder-answer rows.

This audit changes the recovery question but does not hide builder failures.
Across artifact-present builds, builder-side implementation coverage under the
released labels ranges from 70.0\% to 100.0\%, with four builders at 100\%
coverage; Appendix~\ref{app:artifact_audit} reports coverage under adjudicated
labels. Artifact quality is therefore interpreted through implementation
coverage, recovery, stability, and effort jointly.

Task specifications and question banks were schema-validated and manually
reviewed, and the 47 generated artifacts with source files were audited for direct
answer leakage;
we found no such patterns. Appendix~\ref{app:artifact-layout-sensitivity}
decomposes compile-probe exclusions, and Appendix~\ref{app:artifact_audit}
reports the audit scope and qualitative artifact patterns.

\subsection{Prior and stress calibration}
\label{sec:prior_stress}

Table~\ref{tab:prior-stress} calibrates priors, specification access,
artifact-conditioned recovery, and compile-probe exclusions. Question-only
accuracy is 97.2\%, and specification access adds $+2.7$~pp (95\% CI
[$+1.0$,~$+5.0$]). Compile-pass recovery over all answer rows is 96.7\%; on matched
strata it differs from the question-only control by $-0.4$~pp ([$-3.2$,~$+2.6$]). In this
task pack, accuracy with artifact access is therefore indistinguishable from the
question-only baseline, which is the pack's central limitation. In an additional
option-free check on eight questions and two artifacts, free-text answers were
correct in 38 of 48 responses with artifact access and 13 of 48 without it
(Appendix~\ref{app:option_free}). Accuracy serves
as a gate, and the artifact comparison rests on gated inspection effort. Effort
and affinity are interpreted under this calibration and restricted to
compile-passing artifacts.

\begin{table}[t]
\centering
\footnotesize
\setlength{\tabcolsep}{4.0pt}
\caption{Prior, artifact, and stress calibration. All-row accuracy scores every
answer row; $\Delta$ is its difference from question-only on matched finder, task,
and trial strata, with a 95\% question-cluster bootstrap interval
(Appendix~\ref{app:metric_details}), and can differ slightly from the difference
of the displayed rates. Question-only excludes 11 runs (Section~\ref{sec:setup}).
Implemented-gold (impl.-gold) columns use the audited scoring set, carry no
$\Delta$, and omit the 36 runs on the build without source files, which is labeled
absent.}
\label{tab:prior-stress}
\begin{tabular}{lrrrcrr}
\toprule
 & & & \multicolumn{2}{c}{All rows} & \multicolumn{2}{c}{Impl.-gold} \\
\cmidrule(lr){4-5}\cmidrule(lr){6-7}
Slice & Build & Find & Accuracy & \shortstack{$\Delta$ vs.\ question-only\\(pp, 95\% CI)} & Accuracy & Perfect \\
\midrule
Question-only & -- & 61 & 97.2\% & -- & -- & -- \\
Spec-only & -- & 72 & 99.9\% & $+2.7$ [$+1.0$, $+5.0$] & -- & -- \\
Artifact-present, all & 48 & 1728 & 96.4\% & $-0.7$ [$-3.7$, $+2.3$] & 98.9\% & 86.5\% \\
Compile-pass primary & 41 & 1476 & 96.7\% & $-0.4$ [$-3.2$, $+2.6$] & 98.9\% & 86.4\% \\
Compile-failed excluded & 7 & 252 & 94.4\% & $-2.7$ [$-7.2$, $+1.3$] & 98.8\% & 86.6\% \\
\bottomrule
\end{tabular}
\end{table}

\subsection{Stability and repeatability}
\label{sec:stability}

Repeated trials estimate reliability across attempts. Table~\ref{tab:scope}
reports repeatability controls for the formal artifact-conditioned condition and
the question-only and spec-only controls. Low-effort but unstable artifacts
therefore receive weaker evidence than artifacts that are both recoverable and
repeatable.

\begin{table}[t]
\centering
\footnotesize
\setlength{\tabcolsep}{4.0pt}
\caption{Repeatability controls. Formal rows use the audited scoring set;
agreement requires all attempts in a matched cell to be correct (three, or those
remaining where a question-only run is excluded). Reliable questions reach at
least 90\% agreement.}
\label{tab:scope}
\begin{tabular*}{0.80\textwidth}{@{\extracolsep{\fill}}lrrrrr}
\toprule
Slice & Build & Find & Answer rows & \shortstack{3-trial\\agreement} & \shortstack{Reliable\\questions} \\
\midrule
Formal, all tasks & 41 & 1476 & 21312 & 97.5\% & 29 / 30 \\
Spec-only control & -- & 72 & 1080 & 99.7\% & 30 / 30 \\
Question-only control & -- & 61 & 915 & 94.4\% & 24 / 30 \\
\bottomrule
\end{tabular*}
\end{table}

\subsection{Implementation-aware recovery gates}
\label{sec:recovery}

Before interpreting effort, we combine builder-side implementation coverage
with downstream recovery on implemented-gold claims. This prevents an artifact
from looking strong merely because unimplemented claims were excluded from
finder scoring. Implementation-aware downstream recovery ranges from 68.2\% to
99.4\% across builders. Because audited-row downstream recovery is uniformly
high (97.3--99.5\%), most variation comes from builder-side implementation
coverage (70.0--100.0\%), not from finder failure on implemented claims.
Appendix~\ref{app:accuracy_views} gives the full builder-level view.

Table~\ref{tab:recovery} reports the complementary task-level recovery gate.
Entries are computed before inspection-effort normalization or task-length
compounding, reserving effort analysis for Figure~\ref{fig:builderR}. The table
separates the two task families because aggregate recovery can hide
task-specific gaps. Among reported builder--task entries with implemented-gold
evidence, audited answer recovery is already high: every reported entry exceeds
96\%, and most are near or above 98\%. This near-saturation motivates the
conditional inspection-effort analysis in Section~\ref{sec:Rb}. The
parenthesized cell count identifies whether the value rests on the full finder
panel for that task.

\begin{table}[t]
\centering
\caption{Task-level recovery gate. Entries report mean audited answer recovery;
parentheses give contributing finder cells.}
\label{tab:recovery}
\begin{minipage}[t]{0.49\textwidth}
\centering
\textbf{High effort}\\[0.25em]
\footnotesize
\setlength{\tabcolsep}{4.0pt}
\begin{tabular}{@{}lcc@{}}
\toprule
Builder & \texttt{minidb} & \texttt{nanoweb} \\
\midrule
GPT-5.5 & 97.8\% (6/6) & 99.8\% (6/6) \\
GPT-5.4-mini & 98.1\% (6/6) & 100.0\% (6/6) \\
Opus 4.7 & 99.8\% (6/6) & 99.4\% (6/6) \\
Sonnet 4.6 & 98.9\% (6/6) & 99.8\% (6/6) \\
MiMo 2.5 Pro & 99.8\% (6/6) & -- (0/6) \\
MiMo 2.5 & 96.1\% (6/6) & 98.0\% (6/6) \\
\bottomrule
\end{tabular}
\end{minipage}\hfill
\begin{minipage}[t]{0.49\textwidth}
\centering
\textbf{Low effort}\\[0.25em]
\footnotesize
\setlength{\tabcolsep}{4.0pt}
\begin{tabular}{@{}lcc@{}}
\toprule
Builder & \texttt{minidb} & \texttt{nanoweb} \\
\midrule
GPT-5.5 & 98.7\% (6/6) & 99.6\% (6/6) \\
GPT-5.4-mini & 97.6\% (6/6) & 99.4\% (6/6) \\
Opus 4.7 & 99.4\% (6/6) & 99.3\% (6/6) \\
Sonnet 4.6 & 99.4\% (6/6) & 98.8\% (6/6) \\
MiMo 2.5 Pro & -- (0/6) & 98.9\% (6/6) \\
MiMo 2.5 & 98.3\% (6/6) & 97.6\% (6/6) \\
\bottomrule
\end{tabular}
\end{minipage}

\end{table}

\subsection{Conditional inspection effort}
\label{sec:Rb}

\begin{figure}[t]
\centering
\includegraphics[width=\textwidth]{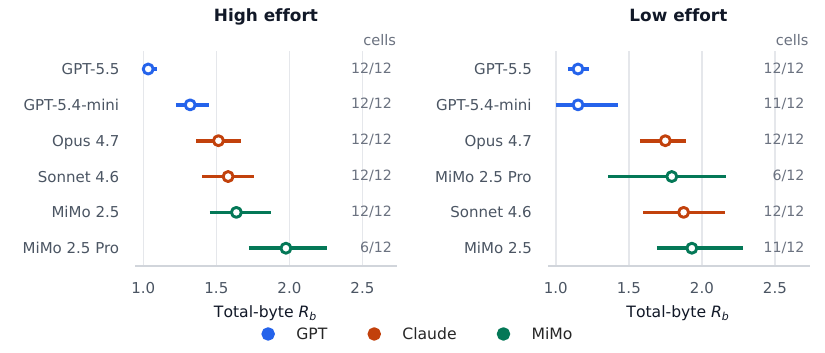}
\caption{Conditional inspection effort, $R_b$, on all-correct cells. Scores use
the audited scoring set; missing cells are failure signals. Bars are 95\%
bootstrap intervals (Appendix~\ref{app:metric_details}).}
\label{fig:builderR}
\end{figure}

With recovery near saturation, inspection effort still reveals differences
among generated artifacts (Figure~\ref{fig:builderR}). The builder effort score
$R_b$ uses cells where a finder answered all questions correctly at least once.
High- and low-effort panels are computed separately, keeping each finder-task
cell's denominator within one effort regime.

Under this metric in the compile-pass panel, the GPT-5.5 rows show low-effort
artifact recovery on all 12 finder--task cells: $R_b=1.033$ in the high-effort
panel and $R_b=1.151$ for GPT-5.5-low in the low-effort panel. GPT-5.4-mini-high
remains second in the high-effort panel ($R_b=1.320$). In the low-effort panel,
GPT-5.4-mini-low is nearly tied with GPT-5.5-low after rounding while
contributing 11/12 all-correct cells; the bootstrap does not separate the two,
and their order depends on the implementation-label set
(Appendix~\ref{app:artifact_audit}, Table~\ref{tab:adjudicated_labels}). This
is a panel-local artifact comparison: artifacts are compared by how much effort
the same finders needed to recover intent in the same finder-task context.
Appendix~\ref{app:usage} reports vendor token usage separately.

Coverage remains part of the interpretation. GPT-5.4-mini-low improves from
7/12 to 11/12 contributing low-effort cells after the audit, but still has one
missing all-correct cell; the task-level recovery gate in
Table~\ref{tab:recovery} keeps such partial-recovery behavior visible.

\subsection{Rank stability}
\label{sec:F2}

The effort score aggregates over finders, so the interpretation depends on
whether finders agree about relative artifact effort. Appendix
Figure~\ref{fig:rankstab} reports Kendall $\tau$ across finder pairs on
total-byte orderings in all-correct cells. Mean $\tau$ over finder pairs is
moderate and positive in the high-effort panel on both tasks (0.49 on
\texttt{scratch\_minidb}, 0.61 on \texttt{scratch\_nanoweb}). Low-effort agreement
is task-dependent: weaker on \texttt{scratch\_minidb} (mean $\tau$ 0.44), stronger
on \texttt{scratch\_nanoweb} (0.55). These
diagnostics guide construction of family-diverse finder panels for follow-on
evaluations.

\subsection{Builder--finder affinity}
\label{sec:affinity_main}

If generated repositories are communication artifacts for future agents, their
legibility may depend on who reads them. The affinity residuals ask whether a
builder--finder pair recovers intent better or worse than expected after
accounting for builder and finder marginal effects. The full matrix in Appendix
Figure~\ref{fig:affinity_residual} reports pair-specific residuals. At the
family level, same-family residuals are positive for OpenAI/Codex (+0.076),
Claude (+0.041), and MiMo (+0.027), while most cross-family cells are near zero
or negative. Appendix~\ref{app:affinity} reports the full 12-agent matrix and
family-level table.

\subsection{Low-prior and evidence-audit sensitivity}
\label{sec:low_prior_main}

Low-prior subsets and artifact-evidence audits test whether recovery depends
on artifact evidence when generic priors are less helpful. The low-prior
subset contains ten questions, five per task, selected on the released controls
by question-only three-trial agreement below 90\%; six of them still meet this
criterion after the exclusion in Section~\ref{sec:setup}. On this subset,
question-only accuracy is 91.8\% and compile-pass recovery over all answer rows is
95.3\%; on matched strata, the difference is $+3.8$~pp ([$-2.6$,~$+10.6$]).
Recovery on the audited low-prior scoring set is 97.9\%.

Table~\ref{tab:low-prior-evidence-summary} summarizes the prior-sensitivity
slice and the post-hoc evidence audit in the main text.

\begin{table}[t]
\centering
\footnotesize
\caption{Low-prior recovery and evidence-audit summary. Question-only excludes
11 runs; recovery scores all compile-pass answer rows, and $\Delta$ is computed as
in Table~\ref{tab:prior-stress}. Impl.-gold recovery uses the audited scoring set.
Audit rows report supported sampled correct answers.}
\label{tab:low-prior-evidence-summary}
\begin{tabular*}{0.90\textwidth}{@{\extracolsep{\fill}}lrrrr}
\toprule
Slice & Question-only & Recovery & $\Delta$ (pp, 95\% CI) & Impl.-gold recovery \\
\midrule
All questions & 97.2\% & 96.7\% & $-0.4$ [$-3.2$, $+2.6$] & 98.9\% \\
Low-prior subset & 91.8\% & 95.3\% & $+3.8$ [$-2.6$, $+10.6$] & 97.9\% \\
\bottomrule
\end{tabular*}

\vspace{0.35em}
\begin{tabular*}{0.90\textwidth}{@{\extracolsep{\fill}}lr}
\toprule
Audit slice & Supported / sampled \\
\midrule
All sampled correct answers & 65 / 72 \\
Low-prior sampled answers & 35 / 36 \\
\bottomrule
\end{tabular*}
\end{table}

The post-hoc artifact-evidence audit samples 72 correct compile-pass answers
and checks whether cited files, symbols, and rationales distinguish the selected
option from alternatives. The audit provides supporting construct evidence:
exact-match recovery in the released records usually has artifact-grounded
support.
In the supported or partially supported audit rows, cited evidence is dominated
by source files (66 of 68 rows), with README/docs or configuration/Cargo files
appearing in 7 rows each and tests absent as the primary cited channel.
Restricted-view ablations such as README-only, source-only, tests-only, or
config-only views require new finder runs and should be reported as a separate
evidence layer. Appendix~\ref{app:low_prior_subset} gives the full low-prior
effort, recovery, rank-stability, and affinity diagnostics, and
Appendix~\ref{app:evidence-audit} gives the sampling rule and labels.

\section{Discussion}

The results show why inspection effort is useful when recovery accuracy
approaches saturation: artifacts that allow the same intended choices to be
recovered can still require different amounts of inspection by the same finder.
The two-stage evaluation uses recovery and stability to establish the basis for
interpreting these effort differences as a proxy for artifact legibility.
Controls quantify prior answerability, while audits expose builder-side
implementation coverage.

Effort matters only because the protocol first asks whether recovery is correct
and repeatable. A repository that is quick to inspect but unstable is not treated
as better than a slower but reliable repository. Conversely, among artifacts
that expose the same intended decisions, the extra inspection required by the
same finder is evidence that the artifact is harder for downstream agents to use
as working context. Affinity diagnostics add a second caution: legibility can be
reader-dependent, so builder rankings should be interpreted together with finder
calibration and family-diverse panels.

\section{Scope and limitations}
\label{sec:validity-scope}

The empirical claims are scoped to the released panel, two task families, finder
policies, and strong priors: question-only accuracy reaches 97.2\%, so accuracy
is a gate while conditional effort carries the artifact-legibility comparison.
Builder orderings are panel-local and conditional on the implementation labels.
The scoring audit is designed to avoid penalizing finders for unimplemented gold
claims, but
\bnf{} measures agent-facing recovery of specification-traced labels. It does
not measure runtime correctness, security, deployment readiness, or human
maintainability. The implementation audit inspects artifacts statically, and the
builder-written test suites pass in 40 of 41 compile-pass builds
(Appendix~\ref{app:artifact-layout-sensitivity}), which indicates
self-consistency; specification-level behavioral testing remains a separate
evaluation. New task packs can use LLM-drafted items when admission rules are
fixed before any builder is evaluated, including question-only and option-free
prior screening and human review of gold answers
(Appendix~\ref{app:benchcard}). Public task packs may also become
contaminated, so private packs remain necessary.

\section{Release, safety, and maintenance}
\label{sec:release}

This study uses synthetic software tasks and no human-subject data. Coding-agent
runs should execute in OS-level sandboxes outside the harness repository, with
resource limits; the released runs used per-run workspaces without such a sandbox
(Appendix~\ref{app:impl}). We release the
harness, task specs, question banks, artifacts, records, canonical tables,
reports, scripts, a reproducibility index, Croissant-compatible metadata
\citep{croissant2024}, licenses, and evidence audit. The code is available at
\url{https://github.com/JacobLinCool/Build-and-Find} under the Apache License 2.0,
and the data at \url{https://huggingface.co/datasets/JacobLinCool/Build-and-Find}
(DOI: 10.57967/hf/11347) under CC BY 4.0. Stored records and
deterministic transforms reproduce the reported analysis; live reruns depend on
provider models, CLIs, endpoints, and drift.
Prior benchmark assets are cited for context and are not redistributed. Model
APIs and CLIs listed in Tables~\ref{tab:agents} and~\ref{tab:execution_stack}
were accessed as provider services under their applicable terms; no model
weights or provider CLI code are redistributed.

\section{Conclusion}

\bnf{} treats generated repositories as communication artifacts for future
agents and evaluates whether downstream agents can efficiently recover their
intended behavior and design choices. Its two-stage evaluation first assesses
recovery and stability, then compares inspection effort. In our task pack, where
several frontier models approach saturation in intent recovery, effort reveals
differences among their generated artifacts within the evaluated finder-task
panel, adding a measure of agent-facing artifact legibility alongside behavioral
correctness.

\begin{ack}
This work received no third-party funding. The author declares no competing
interests.
\end{ack}

\bibliographystyle{plainnat}
\bibliography{references}

\appendix
\raggedbottom

The appendices follow the paper's dependency order: protocol and implementation
substrate; measurement calibration; validity audits; empirical diagnostics; and
release documentation.

\section{Benchmark procedure}
\label{app:procedure}

This appendix states the procedure as pseudocode. It complements the
artifact-facing description in Section~\ref{sec:design} and the harness notes
in Appendix~\ref{app:impl}. Notation follows Section~\ref{sec:metrics}: $b$ is
a builder, $f$ a finder, $t$ a task, and $r$ a trial index. $A_{b,t,r}$ is the
artifact a builder produces, $Q_t$ the question bank, and $y_{t,q}$ the gold
option for question $q$. We write $R_{\mathrm{build}}$ and $R_{\mathrm{find}}$
for per-cell trial counts.

Three properties are surfaced explicitly. First, the specification $s_t$ is
withheld from finders: the finder prompt and workspace contain only the artifact
and the question bank. Second, every invocation is keyed by a base seed
$\mathrm{seed}_0$ plus the trial index, so option order and other harness-side
randomization are reproducible; provider-side sampling is outside this record
(Appendix~\ref{app:details}). Third, find runs are gated on a successful build
invocation: every build with $\mathrm{status}=\texttt{ok}$ produced an artifact
directory and receives artifact-conditioned find runs. The compile probe then
selects the primary analysis panel: artifacts that fail it, and their find
records, are retained for validity-stress diagnostics. Only
agent-side failures (timeouts, CLI errors, infrastructure failures) keep an
artifact out of the canonical artifact-present release.

\begin{algorithm}[H]
\caption{\bnf{} benchmark execution.}
\label{alg:bench}
\begin{algorithmic}[1]
\Require Tasks $\mathcal{T}$ (with hidden specs $s_t$ and banks $Q_t$),
  builders $\mathcal{B}$, finders $\mathcal{F}$, trial counts
  $R_{\mathrm{build}}, R_{\mathrm{find}}$, base seed $\mathrm{seed}_0$.
\Ensure Build records, find records, and the primary analysis panel.
\State $\mathcal{R}_{\mathrm{build}} \gets \emptyset$;\quad
       $\mathcal{R}_{\mathrm{find}} \gets \emptyset$
\ForAll{$t \in \mathcal{T},\ b \in \mathcal{B},\ r \in \{0,\dots,R_{\mathrm{build}}-1\}$}
  \State $\rho \gets \Call{Builder.Run}{b, s_t,\ \mathrm{seed}_0+r,\ r}$
  \State $\mathcal{R}_{\mathrm{build}} \gets \mathcal{R}_{\mathrm{build}} \cup \{\rho\}$
\EndFor
\ForAll{$\rho \in \mathcal{R}_{\mathrm{build}}$ \textbf{with}
        $\rho.\mathrm{status}=\texttt{ok}$,\
        $f \in \mathcal{F}$,\
        $r' \in \{0,\dots,R_{\mathrm{find}}-1\}$}
  \State $\phi \gets \Call{Finder.Run}{f,\ \rho,\ Q_{\rho.t},\
            \mathrm{seed}_0+r',\ r'}$
  \State $\mathcal{R}_{\mathrm{find}} \gets \mathcal{R}_{\mathrm{find}} \cup \{\phi\}$
\EndFor
\State $\mathcal{R}^{\star}_{\mathrm{find}} \gets
       \{\phi \in \mathcal{R}_{\mathrm{find}} : \phi.\rho.\mathrm{compile\_pass}\}$
  \Comment{primary analysis panel}
\State \Return $\mathcal{R}_{\mathrm{build}}$, $\mathcal{R}_{\mathrm{find}}$, $\mathcal{R}^{\star}_{\mathrm{find}}$
\end{algorithmic}
\end{algorithm}

\paragraph{Reported instantiation.}
The reported panel uses $|\mathcal{T}|=2$ tasks (\texttt{scratch\_minidb},
\texttt{scratch\_nanoweb}), $|\mathcal{B}|=|\mathcal{F}|=12$ agent
configurations (Section~\ref{sec:empirical}, Table~\ref{tab:agents}),
$R_{\mathrm{build}}=2$ build trials per $(b,t)$ cell, and
$R_{\mathrm{find}}=3$ find trials per $(\rho,f)$ cell. The build grid contains
$|\mathcal{T}|\cdot|\mathcal{B}|\cdot R_{\mathrm{build}} = 48$ status-ok
artifact-present build runs. Seven fail the release-mode compile probe, leaving
41 compile-pass primary artifacts. Each
task carries $|Q_t|=15$ questions, yielding
$41\cdot|\mathcal{F}|\cdot R_{\mathrm{find}}=1476$ primary
artifact-conditioned find records and $1476\cdot 15 = 22140$ raw question rows.
After this audit, 21312 primary finder-answer rows remain in the scored
recovery set. The spec-only and question-only controls
(Section~\ref{sec:controls}) add $72+72$ control find records under the same
$R_{\mathrm{find}}$. The base seed is $\mathrm{seed}_0=20260501$.

\begin{algorithm}[H]
\caption{Builder trial: synthesize artifact $A_{b,t,r}$.}
\label{alg:builder}
\begin{algorithmic}[1]
\Require Builder $b$, hidden spec $s_t$, seed, trial index $r$.
\Ensure Build record $\rho$.
\State $W \gets \Call{NewWorkspace}{b,t,r}$
  \Comment{fresh per-run directory; seeded with edit-task starter if any}
\State $P_b \gets \Call{RenderBuilderPrompt}{s_t}$
  \Comment{deterministic; $P_b$ and $s_t$ are withheld from finders}
\State $u \gets \Call{Invoke}{b,\ P_b,\ W,\ \tau=s_t.\tau_{\mathrm{build}}}$
  \Comment{records token usage and novel inspection bytes}
\State $A_{b,t,r} \gets W$
  \Comment{the codebase produced inside $W$}
\State $v \gets \Call{Validate}{A_{b,t,r}}$
  \Comment{compile probe and structural metadata}
\State $\rho \gets \langle b,\ t,\ r,\ A_{b,t,r},\ u,\ v,\
            \mathrm{status}=u.\mathrm{status},\
            \mathrm{hash}(s_t),\ \mathrm{hash}(P_b),\ \mathrm{seed} \rangle$
\State \Return $\rho$
\end{algorithmic}
\end{algorithm}

The build record's $\mathrm{status}$ field is taken from the agent invocation
result, so it reflects whether the agent finished cleanly within its time
budget. The compile-probe outcome lives in the validation payload $v$ and is
decoupled from $\mathrm{status}$, so a status-ok artifact may still be retained
as diagnostic evidence even when it is excluded from the primary find panel.

\begin{algorithm}[H]
\caption{Finder trial: recover intent from artifact.}
\label{alg:finder}
\begin{algorithmic}[1]
\Require Finder $f$, build record $\rho$ with artifact $A_{\rho.b,\rho.t,\rho.r}$,
         question bank $Q_t$, seed, trial index $r'$.
\Ensure Find record $\phi$ with per-question correctness.
\State \textbf{require} $\rho.\mathrm{status}=\texttt{ok}$
  \Comment{$\rho.\mathrm{compile\_pass}$ selects the primary panel (Alg.~\ref{alg:bench})}
\State $W' \gets \Call{CloneWorkspace}{A_{\rho.b,\rho.t,\rho.r}}$
  \Comment{fresh writable copy without build outputs or VCS metadata}
\ForAll{$q \in Q_t$}
  \State $\sigma_{r',q} \gets \Call{ShuffleOptions}{q,\
            \mathrm{seed},\ \mathrm{hash}(q)}$
\EndFor
\State $P_f \gets \Call{RenderFinderPrompt}{Q_t,\ \{\sigma_{r',q}\}_q}$
\State $u \gets \Call{Invoke}{f,\ P_f,\ W',\ \tau=s_t.\tau_{\mathrm{find}}}$
\State $\hat{Y} \gets \Call{ReadAnswers}{W'/\texttt{find\_answers.json}}$
\ForAll{$q \in Q_t$}
  \State $\hat{y}_q \gets \sigma_{r',q}^{-1}(\hat{Y}[q])$
    \Comment{de-shuffle to canonical option key}
  \State $\mathrm{correct}_q \gets \mathbf{1}\{\hat{y}_q = y_{t,q}\}$
\EndFor
\State $\phi \gets \langle f,\ \rho,\ r',\
            \{(\hat{y}_q,\mathrm{correct}_q)\}_{q\in Q_t},\ u,\
            \mathrm{seed},\ \mathrm{hash}(P_f) \rangle$
\State \Return $\phi$
\end{algorithmic}
\end{algorithm}

The orchestrator drains the build queue before any find work starts. Records
on disk are reused when their spec hash and prompt hash match the current
configuration; mismatches abort to preserve raw evidence across re-runs.
Per-run usage $u$ is the source of the
inspection-effort metric in Section~\ref{sec:metrics}: \textsc{Invoke} routes
each agent CLI through a local observation layer that records novel retrieval
and tool-call bytes alongside vendor-reported tokens.

\paragraph{Example item.}
Table~\ref{tab:example_item} reproduces item \texttt{q\_nw\_04} of the
\texttt{scratch\_nanoweb} question bank with its specification trace. In one
compile-pass artifact built by GPT-5.4-mini-high,
\texttt{build\_middlewares} in \texttt{src/server.rs} collects the configured
middleware into a \texttt{Vec<Arc<dyn Middleware>>} in configuration order, and
\texttt{dispatch\_route} passes that list to \texttt{middleware::execute}, which
calls the middleware in list order, outermost first, before the route handler.
The pair is implemented-gold under the released and adjudicated labels.

\begin{table}[H]
\centering
\small
\caption{Item \texttt{scratch\_nanoweb:q\_nw\_04} (design-intention recognition).
Letters are canonical; each run presents the options in a seeded shuffled order.}
\label{tab:example_item}
\begin{tabular}{@{}lp{0.82\textwidth}@{}}
\toprule
Question & How is the middleware stack composed? \\
\midrule
A & each middleware is a closure stored in a \texttt{Vec}; the chain is executed
by a fold over that \texttt{Vec} each request \\
B (gold) & each middleware implements a single object-safe trait; the chain is a
list of trait objects whose registration order drives outer-to-inner call order \\
C & middlewares are chosen per request by a proc-macro that expanded at compile
time to a different \texttt{fn} for each route \\
D & middlewares are composed via \texttt{tower::Layer} and
\texttt{tower::Service}; nanoweb re-exports the tower types \\
\midrule
Trace & Structural requirement SR4 (intention I-SR4): ``Middleware is composed
via a single \texttt{Middleware} trait. The trait's object-safe form is what goes
into the per-route chain; the concrete registration order is preserved and
drives the outer-to-inner call order.'' \\
\bottomrule
\end{tabular}
\end{table}

\section{Implementation details}
\label{app:impl}

The harness validates YAML specifications and question banks, renders
deterministic builder and finder prompts, invokes agent CLIs through adapter
registries, and writes append-only run records. Each record contains the task
ID, status, validation or answer payload, and usage metrics where applicable;
its manifest binds the registry agent ID, adapter family, model identifier,
seed, trial ID, prompt hash, specification hash, creation time, \bnf{} version,
and environment hash. A local observation layer records novel retrieval bytes
and novel tool-call bytes at the conversation level, excluding transport
framing, tool schemas, system prompts, and replayed context.

\paragraph{Execution environment.}
Builder and finder workspaces were per-run directories inside the harness
repository, and the agent CLIs ran without an OS-level sandbox. A transcript check
of the 1{,}728 artifact-conditioned find runs found that no primary-panel finder
read the specification or the gold-annotated question files; one run's
search of the parent finder-workspace directory returned other finders' answer files. Excluding the 62
primary-panel runs whose transcripts show reads, file listings, or git commands
that reach outside their workspace, including that run, changes accuracy by at
most 0.11 pp and leaves every $R_b$ rank unchanged.

\section{Additional experimental details}
\label{app:details}

Each task has 15 multiple-choice questions. The reported logical view selects one
successful row per planned logical cell and excludes older pilots and
preserved infrastructure failures. Build trials use planned seed labels 20260501
and 20260502 as run identifiers. Finder trials use planned seeds 20260501,
20260502, and 20260503 both as run identifiers and as deterministic
option-shuffle seeds. When a preserved infrastructure failure requires a rerun,
the harness writes a distinct append-only recovery-seed record and the logical
view selects the successful replacement. These seeds serve as run identifiers
and deterministic shuffle seeds; provider-side sampling seeds are outside this
record. Controls use the same finder panel and trial count and provide either the
specification or the question bank in place of the artifact.
Replacement records preserve the logical trial index; finder option shuffling
uses the recorded recovery seed.

\paragraph{Question-only control exclusions.}
The question-only control excludes runs whose transcripts show reads outside their
workspace: 11 of the 72 runs, of which three read answer-bearing files (the
specification, a gold-annotated question file, or another finder's answer file),
six read the benchmark's seed projects for edit tasks, and two read other runs'
artifact copies. The remaining 61 runs (915 answer rows) give 97.2\% question-only
accuracy; excluding only the three answer-bearing runs gives 96.6\%. The release
lists the excluded runs.

\begin{table}[H]
\centering
\footnotesize
\caption{Logical panel accounting for the reported experiment.
Question rows count graded multiple-choice answers.}
\label{tab:run_accounting}
\begin{tabular}{llrr}
\toprule
Record type & Logical units & Records & Question rows \\
\midrule
Planned build artifacts & 2 tasks $\times$ 12 builders $\times$ 2 trials & 48 & -- \\
Artifact-present generated artifacts & status-ok release path exists & 48 & -- \\
Compile-pass primary artifacts & primary analysis gate & 41 & -- \\
Compile-probe-failed artifacts & excluded diagnostic subset & 7 & -- \\
Formal find runs & 41 artifacts $\times$ 12 finders $\times$ 3 trials & 1{,}476 & 22{,}140 \\
Compile-failed diagnostic find runs & 7 artifacts $\times$ 12 finders $\times$ 3 trials & 252 & 3{,}780 \\
Spec-only controls & 2 tasks $\times$ 12 finders $\times$ 3 trials & 72 & 1{,}080 \\
Question-only controls & 2 tasks $\times$ 12 finders $\times$ 3 trials & 72 & 1{,}080 \\
\bottomrule
\end{tabular}
\end{table}

\section{Agent and harness configuration}
\label{app:reproducibility}

\begin{table}[H]
\centering
\scriptsize
\caption{Agent configurations in the final panel.}
\label{tab:agents}
\begin{tabular}{llll}
\toprule
Label & Harness / provider & Model identifier & Effort \\
\midrule
Claude Opus 4.7 & Claude Code / Anthropic & \texttt{claude-opus-4-7} & high \\
Claude Opus 4.7-low & Claude Code / Anthropic & \texttt{claude-opus-4-7} & low \\
Claude Sonnet 4.6 & Claude Code / Anthropic & \texttt{claude-sonnet-4-6} & high \\
Claude Sonnet 4.6-low & Claude Code / Anthropic & \texttt{claude-sonnet-4-6} & low \\
GPT-5.5 & Codex / OpenAI & \texttt{gpt-5.5} & high \\
GPT-5.5-low & Codex / OpenAI & \texttt{gpt-5.5} & low \\
GPT-5.4 Mini & Codex / OpenAI & \texttt{gpt-5.4-mini} & high \\
GPT-5.4 Mini-low & Codex / OpenAI & \texttt{gpt-5.4-mini} & low \\
MiMo 2.5 Pro & Claude Code / MiMo & \texttt{mimo-v2.5-pro} & high \\
MiMo 2.5 Pro-low & Claude Code / MiMo & \texttt{mimo-v2.5-pro} & low \\
MiMo 2.5 & Claude Code / MiMo & \texttt{mimo-v2.5} & high \\
MiMo 2.5-low & Claude Code / MiMo & \texttt{mimo-v2.5} & low \\
\bottomrule
\end{tabular}
\end{table}

\noindent\textbf{Harness note.} MiMo lacked a native coding-agent harness
satisfying the orchestration and logging requirements of this benchmark. In our
implementation, the MiMo API format was compatible with the Claude Code harness
and incompatible with the Codex harness, so MiMo was invoked through the Claude
Code harness. MiMo rows therefore evaluate model behavior together with the
Claude-Code-style harness policy.

\begin{table}[H]
\centering
\small
\caption{Execution stack for the reported harness and analysis
environment.}
\label{tab:execution_stack}
\begin{tabular}{ll}
\toprule
Component & Version or setting \\
\midrule
Codex CLI & \texttt{codex-cli 0.124.0} \\
Claude Code CLI & \texttt{2.1.114 (Claude Code)} \\
\bnf{} harness & \texttt{0.1.0} \\
Python & \texttt{3.13.5} \\
\texttt{uv} & \texttt{0.9.8} \\
Rust toolchain & \texttt{rustc 1.94.0}; \texttt{cargo 1.94.0} \\
Codex effort knob & \texttt{model\_reasoning\_effort=high/low} \\
Claude Code effort knob & \texttt{--effort high/low} \\
\bottomrule
\end{tabular}
\end{table}

Run manifests record the exact registry agent ID, adapter family, model
identifier, prompt hash, specification hash, seed, trial ID, creation time,
\bnf{} version, and environment hash for every build and find invocation. The
release preserves these manifest fields as columns of the derived tables so
that reported numbers can be audited without relying on appendix-level command
transcripts.

The stored records and deterministic transforms reproduce the reported
analysis. Live reruns of builder and finder interactions depend on provider
model availability, CLI versions, provider endpoints, and model drift; reruns
should therefore be treated as new evidence layers with their own records.

\section{Finder calibration}
\label{app:finder_calibration}

Table~\ref{tab:finder_calibration} reports the per-finder calibration evidence
behind the instrument framing in Section~\ref{sec:design}. Formal recovery is
artifact-conditioned exact-match recovery over all compile-pass answer rows. The
$\Delta$ columns report the difference from the same finder's question-only
control over matched task and trial strata: overall, on low-prior questions, and
on implementation-signal questions.
Spec-only controls are saturated in this task pack and are retained as validity
context in the generated calibration summaries.

\begin{table}[H]
\centering
\scriptsize
\setlength{\tabcolsep}{3.4pt}
\caption{Finder calibration summary. Formal is a percentage of all
compile-pass answer rows; $\Delta$ columns are percentage-point differences from
the finder's question-only control after the exclusions of
Section~\ref{sec:setup}. Mean tokens include subagent calls and are per compile-pass find run, in thousands.}
\label{tab:finder_calibration}
\begin{tabular}{lrrrrr}
\toprule
Finder & Formal & Q-only $\Delta$ & Low-prior $\Delta$ & Impl.-signal $\Delta$ & Mean tok. \\
\midrule
Sonnet 4.6-low & 98.7\% & +0.9 & +3.4 & +2.2 & 671 \\
Sonnet 4.6-high & 98.0\% & +2.5 & +10.2 & +5.6 & 729 \\
Opus 4.7-low & 97.9\% & -2.1 & -2.6 & -0.7 & 521 \\
Opus 4.7-high & 97.1\% & -2.9 & -3.1 & -0.8 & 913 \\
MiMo-v2.5-high & 96.8\% & +1.7 & +9.4 & +5.4 & 404 \\
MiMo-v2.5-low & 96.8\% & +2.8 & +12.7 & +7.1 & 398 \\
MiMo-v2.5-pro-low & 96.7\% & +12.9 & +36.7 & +23.0 & 412 \\
MiMo-v2.5-pro-high & 96.7\% & +5.0 & +19.9 & +10.5 & 413 \\
GPT-5.5-low & 96.5\% & -3.5 & -3.4 & -1.0 & 389 \\
GPT-5.5-high & 95.9\% & -4.1 & -4.2 & -1.5 & 547 \\
GPT-5.4-mini-low & 95.2\% & -2.1 & +0.4 & +0.5 & 329 \\
GPT-5.4-mini-high & 94.1\% & -5.9 & -7.3 & -3.6 & 638 \\
\bottomrule
\end{tabular}
\end{table}

The post-calibration core-finder selector in
\texttt{scripts/select\_eandd\_core\_finders.py} applies hard gates for
formal recovery, finder-specific question-only lift, low-prior lift,
implementation-signal lift, cost, rank-stability, and affinity. With token totals
including subagent calls, no configuration satisfies the routine-panel cost
ceiling of 150{,}000 tokens per run. MiMo-v2.5-pro-low remains a prior-stress
audit finder; it has the largest question-only lift in
Table~\ref{tab:finder_calibration} (+12.9~pp). The generated selection artifact
is \texttt{runs\_release/reports/eandd\_final/core\_finder\_selection/}.

\section{Metric details}
\label{app:metric_details}

The reported effort unit is novel agent inspection bytes, the sum of input-side
\texttt{novel\_retrieval\_bytes} and output-side
\texttt{novel\_tool\_call\_bytes}. Retrieval bytes are distinct
\texttt{tool\_result} bytes delivered to the model. Tool-call bytes are emitted
arguments for search and read actions. System prompts, tool schemas, transport
overhead, and replayed context are excluded.

All effort scores are computed inside a fixed analysis panel with builders
$\mathcal{B}$, finders $\mathcal{F}$, and tasks $\mathcal{T}$. For an effort
metric $m$, let $G_{b,f,t}$ be the all-correct runs for builder $b$, finder $f$,
and task $t$ on the audited scoring set. When this set is nonempty, the cell
cost is
\begin{equation}
  c_m(b,f,t)=\frac{1}{|G_{b,f,t}|}\sum_{r\in G_{b,f,t}} m(b,f,t,r).
\end{equation}
If no all-correct run exists, the cell is excluded from the conditional effort
average and counted in coverage. Within a fixed high- or low-effort panel, the
within-finder ratio is
\begin{equation}
  r_m(b,f,t) =
  \frac{c_m(b,f,t)}
       {\min_{b'\in\mathcal{B}:\,c_m(b',f,t)\ \mathrm{defined}} c_m(b',f,t)} .
\end{equation}
Let $\mathcal{C}_b$ be the finder-task cells where builder $b$ has a defined
all-correct cost. The conditional builder-level artifact-effort diagnostic
is
\begin{equation}
  R_m(b)=\exp\left(\frac{1}{|\mathcal{C}_b|}
  \sum_{(f,t)\in\mathcal{C}_b} \log r_m(b,f,t)\right).
\end{equation}
The tables write this score as $R_b$ for the reported total-byte view and report
$|\mathcal{C}_b|$ explicitly.

For the task-level recovery gate, define
\begin{equation}
  g(b,f,t)=\bar{\acc}(b,f,t),
\end{equation}
where $\bar{\acc}$ is computed on the audited scoring set and bars denote means
over repeated find runs. For each builder--task table entry, let
$\mathcal{D}_{b,t}$ be the set of contributing finders for builder $b$ on task
$t$. The table reports
\begin{equation}
  G_{\mathrm{task}}(b,t)=
  \frac{1}{|\mathcal{D}_{b,t}|}\sum_{f\in\mathcal{D}_{b,t}} g(b,f,t),
\end{equation}
with the corresponding contributing finder-cell count reported in parentheses.

\paragraph{Effort estimand.}
$R_b$ is finder-conditioned. Each ratio $r_m(b,f,t)$ compares builders within a
single finder--task cell of one effort panel, and $R_b$ is the geometric mean of
these ratios over the cells where builder $b$ has an all-correct run. It
therefore describes how much inspection a given finder spends on a builder's
artifacts in runs that answer every scored question correctly, relative to the
builder with the lowest such effort for the same finder and task. Because the
ratios depend on each finder's search policy, $R_b$ is panel-local: it describes
artifacts as inspected by the finders of one panel. It is also conditional on the
implementation labels, which fix the questions scored in each cell.
Section~\ref{sec:F2} reports how far finders agree on these orderings.

\paragraph{Answer correctness and evidence sufficiency.}
These are separate views of a finder answer. Answer correctness is the exact
match between the de-shuffled selected option and the specification gold
option; every accuracy, gate, and effort metric uses it. Evidence sufficiency
asks whether the files, symbols, and rationale cited with the answer distinguish
the selected option from the alternatives in the artifact. A correct answer can
rest on insufficient evidence, for example when it follows from generic priors.
Correctness is scored on every answer row, and evidence sufficiency is audited
by sample (Appendix~\ref{app:evidence-audit}); the option-free check grades
correctness and whether the cited files contain the implementing code on every
artifact-condition answer (Appendix~\ref{app:option_free}).

\paragraph{Uncertainty.}
Intervals for differences from the question-only control are 95\% percentile
intervals from 2{,}000 bootstrap resamples of task--question clusters (30 overall,
10 for the low-prior subset, and 15 per task). Both arms use matched finder, task,
and trial strata: artifact-conditioned and spec-only rows leave the contrast when
the question-only run of the same finder, task, and trial is excluded. $R_b$
intervals in Figures~\ref{fig:builderR} and~\ref{fig:low_prior_builderR} are 95\%
percentile intervals from 2{,}000 resamples of the 12 finder--task cells of each
effort panel; repeated draws of a cell are kept as separate cells, and the per-cell
normalization and geometric mean are recomputed in each resample.

\section{Artifact layout sensitivity audit}
\label{app:artifact-layout-sensitivity}

The released primary panel uses a root-level compile probe: an artifact must
present a complete Cargo project at the artifact root and pass
\texttt{cargo build --release}. This gate treats project layout as both a
communication artifact and packaging metadata, because downstream finders
receive the artifact root as the codebase boundary.

To separate layout fidelity from Rust source buildability, we ran a post-hoc
sensitivity audit on the seven artifact-present builds excluded by the primary
compile gate. The audit used existing artifacts without rerunning builders or
modifying outputs. If an excluded artifact had no root \texttt{Cargo.toml} but contained exactly one
nested \texttt{Cargo.toml}, we copied that nested crate to a temporary directory
and ran \texttt{cargo build --release} with an isolated target directory. Six of
the seven excluded builds contained a unique nested crate, and all six built
successfully under this relaxed probe. The remaining excluded build contained no
Rust source files and no nested Cargo project. Because an artifact without source
files implements no specified claim, this build is labeled absent/ambiguous on all
15 of its questions.

\begin{table}[H]
\centering
\small
\caption{Post-hoc sensitivity audit for primary compile-probe exclusions. The
relaxed probe supplies a diagnostic view alongside the primary panel.}
\label{tab:artifact_layout_sensitivity}
\begin{tabular}{lr}
\toprule
Probe slice & Builds \\
\midrule
Primary root-level compile-probe exclusions & 7 \\
Unique nested Cargo crate found & 6 \\
Nested crate builds under relaxed probe & 6 \\
Nested crate build failures & 0 \\
Nested crate build timeouts & 0 \\
No nested Cargo crate found & 1 \\
\bottomrule
\end{tabular}
\end{table}

This audit shows that the excluded set is dominated by artifact-layout
violations: six builds produced buildable source one directory below the
expected artifact boundary, while one build
contained no Rust source files or nested Cargo project. We therefore keep the
primary estimand unchanged, but interpret the compile-probe exclusions as a
mixture of layout-contract failures and no-source output diagnostics.

\paragraph{Builder-written test suites.}
In a post-hoc run, \texttt{cargo build --release --locked} followed by
\texttt{cargo test --locked} was executed in the 41 released artifacts with a
root \texttt{Cargo.toml}. The 41 compile-pass artifacts
built; the other seven have no root \texttt{Cargo.toml} and are the seven
compile-probe exclusions. In 40 of the 41 compile-pass builds the full suite
passed, and across all 41 suites 2{,}060 of 2{,}061 tests passed. The one failing
test is \texttt{cors\_sets\_origin} in a GPT-5.5-low \texttt{scratch\_nanoweb}
build, which passes 25 of its 26 tests. Each builder wrote its own tests, so
these results indicate self-consistency of an artifact with its own suite;
specification-level behavioral testing remains a separate evaluation. The
release reports per-builder and per-task counts.

\section{Artifact audit and qualitative builder patterns}
\label{app:artifact_audit}

We manually inspected the 47 generated build artifacts with source files in the
canonical artifact-present release, which spans two tasks, twelve builders, and
two build trials per builder-task cell. The
audit covered source files, tests, README files, configuration files, and Cargo
metadata. It looked for direct gaming signals: question IDs, answer-option
labels, gold-answer keys, answer-distribution hints, copied question or option
phrasing, benchmark requirement or intention IDs embedded in artifacts, and
documentation that repeated evaluation wording in place of describing the
implementation. We found no evidence of these direct leakage patterns in the
generated artifacts. This audit is a canonical-release validity check;
adaptive-attack robustness remains a separate stress-test setting.

Table~\ref{tab:implementation-coverage} reports builder-side implementation
coverage over the artifact-present release. Coverage is the fraction of
artifact-question pairs audited as implemented-gold. This is a builder-side
diagnostic, separate from finder recovery and conditional inspection effort.

\begin{table}[!htbp]
\centering
\small
\setlength{\tabcolsep}{4.0pt}
\caption{Builder-side implementation coverage over artifact-present builds under
the released labels; the end of this appendix reports coverage under adjudicated
labels. Coverage is the fraction of artifact-question pairs audited as
implemented-gold; the MiMo 2.5-low build without source files is labeled absent.}
\label{tab:implementation-coverage}
\begin{tabular}{lrrrr}
\toprule
Builder & Pairs & Implemented-gold & Coverage & Non-gold / ambig. \\
\midrule
Opus 4.7-high & 60 & 60 & 100.0\% & 0 / 0 \\
Opus 4.7-low & 60 & 60 & 100.0\% & 0 / 0 \\
GPT-5.4-mini-high & 60 & 60 & 100.0\% & 0 / 0 \\
GPT-5.5-high & 60 & 60 & 100.0\% & 0 / 0 \\
GPT-5.5-low & 60 & 59 & 98.3\% & 1 / 0 \\
Sonnet 4.6-high & 60 & 58 & 96.7\% & 2 / 0 \\
Sonnet 4.6-low & 60 & 57 & 95.0\% & 3 / 0 \\
MiMo 2.5 Pro-high & 60 & 56 & 93.3\% & 4 / 0 \\
MiMo 2.5 Pro-low & 60 & 56 & 93.3\% & 4 / 0 \\
MiMo 2.5-high & 60 & 53 & 88.3\% & 4 / 3 \\
GPT-5.4-mini-low & 60 & 51 & 85.0\% & 8 / 1 \\
MiMo 2.5-low & 60 & 42 & 70.0\% & 2 / 16 \\
\bottomrule
\end{tabular}
\end{table}

The same inspection exposed qualitative differences in how builder families
communicate design intent. Codex-built artifacts were generally more concise:
they tended to use flatter crate layouts, shorter README files, and source
organization as the main carrier of design intent. Claude Opus and Sonnet
artifacts more often included longer README architecture sections and richer
module-level comments, making design choices explicit in prose as well as code.
MiMo artifacts were more variable, often with thicker scaffolding and more
extensive project shells, but less consistent structure across trials. These
patterns provide descriptive context for why artifact legibility can interact
with finder family: downstream agents may benefit differently from terse
code-structure signals versus explicit prose documentation.

Implementation labels for the 592 scored compile-pass pairs come from
finder-consensus triage (589 pairs) and manual review (3 pairs). Post-submission
source checks found nine scored pairs whose implemented-gold label the source code
does not support. A model-assisted source trace of all 592 pairs then flagged 205
pairs for review, including all nine. The author adjudicated all 205 flags: the
43 flags on the two low-effort Codex builders (41 relabeled, 2 kept), and the
other 162 by reviewing and approving decisions drafted by model-assisted
adjudication (154 relabeled, 8 kept; two passes per flag, with arbitration for the
7 flags on which the passes disagreed). In total, 195 pairs are relabeled away
from implemented-gold (190 to
absent/ambiguous and 5 to implemented-non-gold), leaving 397 of the 592 scored
pairs. Implemented-gold accuracy on the remaining pairs is 99.7\% overall and
99.4\% on the low-prior subset, compared with 98.9\% and 97.9\% under the
released labels. On the 195 relabeled pairs, finders chose the gold option in
97.3\% of answer rows (6,829 of 7,020). Under the adjudicated labels,
builder-side implementation coverage over artifact-present builds ranges from
35.0\% to 83.3\%, and no builder reaches 100\%.

Three \texttt{scratch\_nanoweb} questions account for 59 of the 195 relabeled
pairs: \texttt{q\_nw\_02}, \texttt{q\_nw\_11}, and \texttt{q\_nw\_14} are
implemented-gold for at most one of the 20 compile-pass artifacts after
adjudication. The gold option of \texttt{q\_nw\_02} includes the per-request
allocation bound of the routing requirement, which almost no artifact meets. The
two negated questions usually admit a second correct option, so their part of
the coverage reduction comes from question wording. The adjudication covers
flagged pairs only. Unflagged pairs were not re-reviewed, and some share the
issue of a relabeled pair; on \texttt{q\_nw\_07}, for example, some unflagged
and some relabeled pairs use the raw request path as the metrics route label.

Table~\ref{tab:adjudicated_labels} compares $R_b$ under the released and
adjudicated labels. The high-effort order is unchanged, the two low-effort
leaders swap, and low-effort ranks three to six keep their order. The swap
already appears when the nine observed errors and the 43 author decisions are
applied, and adding the 162 model-assisted decisions leaves both leaders' $R_b$
unchanged. Primary analyses use the released labels, in which the build
without source files is labeled absent
(Appendix~\ref{app:artifact-layout-sensitivity}).

\begin{table}[!htbp]
\centering
\small
\setlength{\tabcolsep}{4.0pt}
\caption{Conditional inspection effort, $R_b$, under released and adjudicated
implementation labels. Rows follow the released order within each panel. Cells is
the number of contributing all-correct finder--task cells; Q/cell is the mean
number of scored questions per contributing cell.}
\label{tab:adjudicated_labels}
\begin{tabular}{lrrrrrrrr}
\toprule
 & \multicolumn{4}{c}{Released labels} & \multicolumn{4}{c}{Adjudicated labels} \\
\cmidrule(lr){2-5}\cmidrule(lr){6-9}
Builder & Rank & $R_b$ & Cells & Q/cell & Rank & $R_b$ & Cells & Q/cell \\
\midrule
\multicolumn{9}{l}{\emph{High-effort panel}} \\
GPT-5.5-high & 1 & 1.0327 & 12/12 & 15.00 & 1 & 1.0327 & 12/12 & 11.75 \\
GPT-5.4-mini-high & 2 & 1.3200 & 12/12 & 15.00 & 2 & 1.3135 & 12/12 & 11.00 \\
Opus 4.7-high & 3 & 1.5136 & 12/12 & 15.00 & 3 & 1.5016 & 12/12 & 12.00 \\
Sonnet 4.6-high & 4 & 1.5797 & 12/12 & 14.50 & 4 & 1.5575 & 12/12 & 10.25 \\
MiMo 2.5-high & 5 & 1.6364 & 12/12 & 13.25 & 5 & 1.6018 & 12/12 & 5.50 \\
MiMo 2.5 Pro-high & 6 & 1.9757 & 6/12 & 14.00 & 6 & 1.9275 & 6/12 & 9.50 \\
\midrule
\multicolumn{9}{l}{\emph{Low-effort panel}} \\
GPT-5.5-low & 1 & 1.1510 & 12/12 & 14.75 & 2 & 1.1681 & 12/12 & 10.50 \\
GPT-5.4-mini-low & 2 & 1.1514 & 11/12 & 13.45 & 1 & 1.1280 & 12/12 & 5.75 \\
Opus 4.7-low & 3 & 1.7492 & 12/12 & 15.00 & 3 & 1.7368 & 12/12 & 12.50 \\
MiMo 2.5 Pro-low & 4 & 1.7933 & 6/12 & 15.00 & 4 & 1.8358 & 6/12 & 8.00 \\
Sonnet 4.6-low & 5 & 1.8743 & 12/12 & 14.25 & 5 & 1.8934 & 12/12 & 9.75 \\
MiMo 2.5-low & 6 & 1.9306 & 11/12 & 14.00 & 6 & 1.9829 & 12/12 & 6.75 \\
\bottomrule
\end{tabular}
\end{table}

\section{Post-hoc artifact-evidence audit}
\label{app:evidence-audit}

Exact-match grading makes the benchmark reproducible, but exact-match
correctness alone cannot show whether a finder used artifact evidence. To
connect the deterministic score to the intended construct, we audited a
stratified sample of correct finder answers using the structured evidence fields
already present in finder records. The formal score remains exact-match; this
audit checks alignment between exact-match recovery and artifact-grounded
rationales in the released records.

We sampled 72 correct answers from the compile-pass artifact-conditioned panel
with audit seed 20260601, stratifying by task, builder family, finder family,
and low-prior versus high-prior question status. The sample excludes
question-only and spec-only controls. For each sampled answer, the auditor
inspected the recorded rationale, evidence files, evidence symbols, and the
corresponding artifact files.

We assign one of four labels. \emph{Supported} means the cited artifact
evidence directly supports the selected answer and distinguishes it from
plausible alternatives. \emph{Partially supported} means the cited evidence is
relevant while leaving the gold option underdetermined.
\emph{Unsupported} means the cited evidence is absent, irrelevant, or
contradicted by the artifact. \emph{Prior-like} means the rationale mainly uses
generic engineering priors or question wording.

\begin{table}[H]
\centering
\small
\caption{Post-hoc artifact-evidence audit over sampled correct finder answers.
Unsupported and prior-like labels are combined in the final column.}
\label{tab:evidence_audit}
\begin{tabular}{lrrrr}
\toprule
Slice & $N$ & Supported & Partial & Unsupported/Prior-like \\
\midrule
All sampled correct answers & 72 & 65 & 3 & 4 \\
Low-prior questions & 36 & 35 & 0 & 1 \\
\texttt{scratch\_minidb} & 36 & 31 & 2 & 3 \\
\texttt{scratch\_nanoweb} & 36 & 34 & 1 & 1 \\
\bottomrule
\end{tabular}
\end{table}

This audit evaluates construct alignment for the released exact-match records by
checking whether existing correct answers often carry plausible
artifact-grounded rationales. Among the 68 supported or partially supported rows,
cited evidence is concentrated in source files (66 rows); README/docs and
configuration/Cargo files appear in 7 rows each, with tests absent as the primary
cited channel in this sample. Under the adjudicated labels of
Appendix~\ref{app:artifact_audit}, 23 of the 65 supported answers (10 of the 35
low-prior ones) fall on pairs relabeled away from implemented-gold.
Restricted-view ablations such as README-only, source-only, tests-only,
config-only, and source+tests without README views require new live finder runs
and should be released as a separate evidence layer. With independently
validated reference evidence for each pair, a future task pack can require these
evidence fields and grade evidence sufficiency for every scored pair, reported
separately from answer correctness.

\section{Accuracy and robustness views}
\label{app:accuracy_views}

\begin{figure}[H]
\centering
\includegraphics[width=\textwidth]{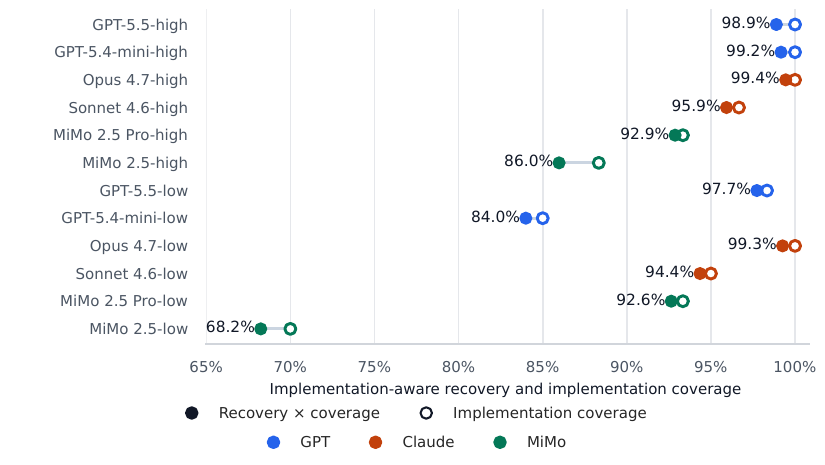}
\caption{Implementation-aware builder downstream recovery diagnostic. Filled
markers multiply downstream recovery on audited rows by builder-side
implementation coverage; open markers show coverage alone. The axis is
truncated, and values are diagnostic rather than a leaderboard.}
\label{fig:builderacc}
\end{figure}

Builder downstream accuracy provides validity context for the artifact-effort
diagnostics. Figure~\ref{fig:builderacc} reports an implementation-aware
recovery view: it combines builder-side implementation coverage with downstream
recovery on audited scoring rows, while the effort views answer how much work
the same finders spend.
For builder $b$, the plotted rate is
\begin{equation}
  D_b =
  \left(
  \frac{\sum_{f,t,a,s}\sum_{q\in Q^+_{b,t,a}}\mathbf{1}\{\hat{y}_{b,f,t,a,s,q}=y_{t,q}\}}
       {\sum_{f,t,a,s}|Q^+_{b,t,a}|}
  \right)
  \left(
  \frac{\sum_{t,a}|Q^+_{b,t,a}|}{\sum_{t,a}|Q_t|}
  \right).
\end{equation}
The first factor is downstream recovery on the audited scoring set; the second
factor is builder-side implementation coverage over artifact-present builds.

\begin{figure}[!htbp]
\centering
\includegraphics[width=\textwidth]{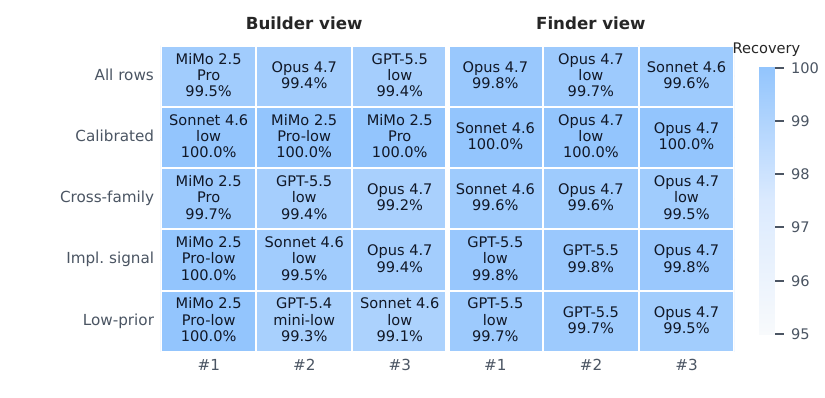}
\caption{Control-conditioned robustness views. Values are exact-match recovery
percentages over audited scoring rows.}
\label{fig:robustviews}
\end{figure}

Figure~\ref{fig:robustviews} reports pivot views that decouple several effects:
which finders are used as instruments, whether builder and finder families are
matched, and whether the questions are high-prior or implementation-sensitive.
These slices are robustness checks around the reported artifact comparison. They
show why the paper uses multiple views: raw recovery, finder accuracy, and
conditional artifact-efficiency emphasize complementary parts of the panel.

Slice definitions are fixed by the analysis script. The calibrated-instrument
row restricts the builder view to the four highest artifact-conditioned recovery
finders and the finder view to the four highest downstream-recovery builders.
Low-prior-answerability questions are those with question-only three-trial
agreement below 90\%. Implementation-signal questions are those where
artifact-conditioned three-trial agreement exceeds question-only agreement by at
least 5 percentage points; both sets are selected on the released question-only
controls (all 72 runs). Cross-family rows exclude same-family builder--finder
pairs.

\begin{table}[H]
\centering
\scriptsize
\setlength{\tabcolsep}{3.0pt}
\caption{Task-specific rows corresponding to Table~\ref{tab:prior-stress}, with
the same columns and conventions; intervals resample the 15 questions of each
task.}
\label{tab:prior-stress-by-task}
\begin{tabular}{llrrrcrr}
\toprule
 & & & & \multicolumn{2}{c}{All rows} & \multicolumn{2}{c}{Impl.-gold} \\
\cmidrule(lr){5-6}\cmidrule(lr){7-8}
Slice & Task & Build & Find & Accuracy & $\Delta$ (pp, 95\% CI) & Accuracy & Perfect \\
\midrule
Question-only & \texttt{scratch\_minidb} & -- & 30 & 98.2\% & -- & -- & -- \\
Question-only & \texttt{scratch\_nanoweb} & -- & 31 & 96.1\% & -- & -- & -- \\
Spec-only & \texttt{scratch\_minidb} & -- & 36 & 99.8\% & $+1.6$ [$+0.2$, $+3.1$] & -- & -- \\
Spec-only & \texttt{scratch\_nanoweb} & -- & 36 & 100.0\% & $+3.9$ [$+0.9$, $+8.0$] & -- & -- \\
Artifact-present, all & \texttt{scratch\_minidb} & 24 & 864 & 95.8\% & $-2.3$ [$-5.0$, $+0.4$] & 98.7\% & 83.2\% \\
Artifact-present, all & \texttt{scratch\_nanoweb} & 24 & 864 & 96.9\% & $+0.8$ [$-4.4$, $+5.8$] & 99.1\% & 89.9\% \\
Compile-pass primary & \texttt{scratch\_minidb} & 21 & 756 & 96.4\% & $-1.7$ [$-4.2$, $+0.8$] & 98.7\% & 82.7\% \\
Compile-pass primary & \texttt{scratch\_nanoweb} & 20 & 720 & 97.0\% & $+0.9$ [$-4.1$, $+6.0$] & 99.2\% & 90.4\% \\
\bottomrule
\end{tabular}
\end{table}

\begin{table}[H]
\centering
\small
\caption{Task-specific repeatability rows omitted from Table~\ref{tab:scope}.}
\label{tab:scope-by-task}
\begin{tabular}{lrrrrr}
\toprule
Slice & Build & Find & Answer rows & \shortstack{3-trial\\agreement} & \shortstack{Reliable\\questions} \\
\midrule
\texttt{scratch\_minidb}, formal & 21 & 756 & 10872 & 96.9\% & 14 / 15 \\
\texttt{scratch\_nanoweb}, formal & 20 & 720 & 10440 & 98.2\% & 15 / 15 \\
\bottomrule
\end{tabular}
\end{table}

Table~\ref{tab:rb_by_task} repeats the $R_b$ computation of
Figure~\ref{fig:builderR} on each task separately. Per-task orders keep the
pooled top two builders in both panels, in reversed order on
\texttt{scratch\_nanoweb} in the low-effort panel, and every pair of builders
whose per-task order departs from the pooled order has overlapping per-task
intervals.

\begin{table}[!htbp]
\centering
\scriptsize
\setlength{\tabcolsep}{4.5pt}
\caption{Per-task conditional inspection effort, $R_b$, under the released
labels. Each task is scored with the rule of Figure~\ref{fig:builderR}
(all-correct gate, total bytes) on its six finder cells per effort panel;
intervals resample those cells as in Appendix~\ref{app:metric_details}. Pooled
is the rank in Figure~\ref{fig:builderR}; -- marks a builder without a
compile-pass artifact on the task.}
\label{tab:rb_by_task}
\begin{tabular}{lrrrcrrrcr}
\toprule
 & & \multicolumn{4}{c}{\texttt{scratch\_minidb}} & \multicolumn{4}{c}{\texttt{scratch\_nanoweb}} \\
\cmidrule(lr){3-6}\cmidrule(lr){7-10}
Builder & Pooled & Rank & $R_b$ & 95\% CI & Cells & Rank & $R_b$ & 95\% CI & Cells \\
\midrule
\multicolumn{10}{l}{\emph{High-effort panel}} \\
GPT-5.5-high & 1 & 1 & 1.061 & [1.000, 1.196] & 6/6 & 1 & 1.005 & [1.000, 1.014] & 6/6 \\
GPT-5.4-mini-high & 2 & 2 & 1.345 & [1.158, 1.608] & 6/6 & 2 & 1.296 & [1.270, 1.326] & 6/6 \\
Opus 4.7-high & 3 & 4 & 1.648 & [1.461, 1.893] & 6/6 & 3 & 1.390 & [1.211, 1.539] & 6/6 \\
Sonnet 4.6-high & 4 & 3 & 1.571 & [1.260, 1.938] & 6/6 & 5 & 1.588 & [1.502, 1.723] & 6/6 \\
MiMo 2.5-high & 5 & 5 & 1.891 & [1.629, 2.276] & 6/6 & 4 & 1.416 & [1.326, 1.561] & 6/6 \\
MiMo 2.5 Pro-high & 6 & 6 & 1.976 & [1.755, 2.230] & 6/6 & -- & -- & -- & -- \\
\midrule
\multicolumn{10}{l}{\emph{Low-effort panel}} \\
GPT-5.5-low & 1 & 1 & 1.068 & [1.014, 1.131] & 6/6 & 2 & 1.240 & [1.147, 1.316] & 6/6 \\
GPT-5.4-mini-low & 2 & 2 & 1.104 & [1.000, 1.345] & 5/6 & 1 & 1.193 & [1.000, 1.696] & 6/6 \\
Opus 4.7-low & 3 & 5 & 1.799 & [1.702, 1.899] & 6/6 & 3 & 1.701 & [1.421, 1.957] & 6/6 \\
MiMo 2.5 Pro-low & 4 & -- & -- & -- & -- & 4 & 1.793 & [1.399, 2.124] & 6/6 \\
Sonnet 4.6-low & 5 & 3 & 1.601 & [1.297, 1.888] & 6/6 & 5 & 2.194 & [1.946, 2.512] & 6/6 \\
MiMo 2.5-low & 6 & 4 & 1.699 & [1.529, 1.891] & 6/6 & 6 & 2.250 & [1.863, 2.935] & 5/6 \\
\bottomrule
\end{tabular}
\end{table}

\begin{figure}[!htbp]
\centering
\includegraphics[width=0.82\textwidth]{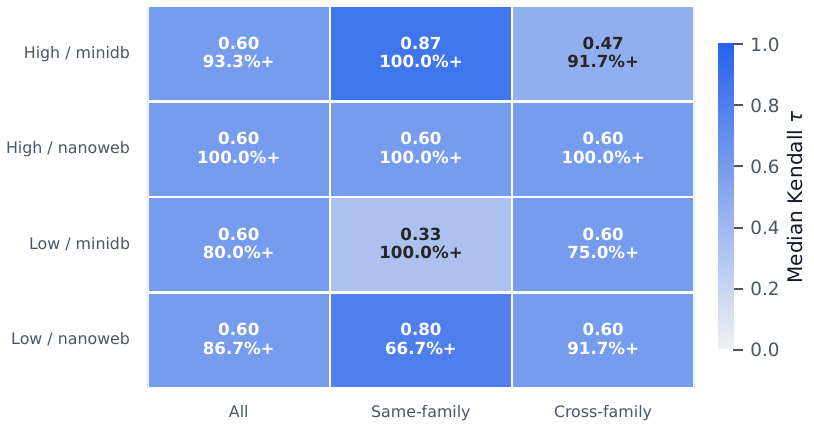}
\caption{Pairwise finder agreement on total-byte builder-effort orderings in
all-correct cells over the audited scoring set: median Kendall $\tau$ over
finder pairs and share of pairs with positive $\tau$ (mean $\tau$ in
Section~\ref{sec:F2}).}
\label{fig:rankstab}
\end{figure}

\FloatBarrier
\section{Affinity diagnostics}
\label{app:affinity}

\begin{figure}[!htbp]
\centering
\includegraphics[width=\textwidth]{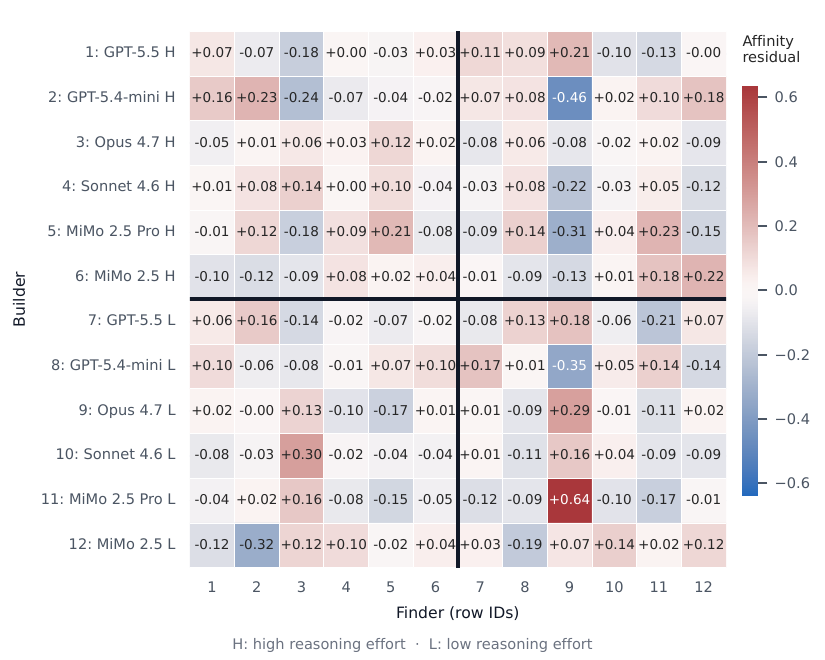}
\caption{Pair-specific builder--finder affinity residuals for the 12-agent
compile-pass panel. Scores use the audited scoring set. Red indicates
above-marginal recoverability; blue indicates below-marginal recoverability.}
\label{fig:affinity_residual}
\end{figure}

Affinity residuals are diagnostic checks for pair-specific builder--finder
effects. They are computed by taking the recovery score used in the matrix view,
subtracting the builder mean and finder mean, and adding back the grand mean.
Positive values indicate above-marginal recovery for a builder--finder pairing in
this panel. These diagnostics have panel-local scope. Figure~\ref{fig:affinity_residual}
shows the full 12-agent matrix, and Table~\ref{tab:family_affinity} aggregates
those residuals by model family.

\begin{table}[!htbp]
\centering
\small
\caption{Family-level affinity residuals in the compile-pass panel. Positive
values are above marginal expectations.}
\label{tab:family_affinity}
\begin{tabular}{lrrr}
\toprule
Builder family & OpenAI/Codex finder & Anthropic/Claude finder & MiMo finder \\
\midrule
OpenAI/Codex & 0.076 & -0.077 & 0.001 \\
Anthropic/Claude & -0.013 & 0.041 & -0.028 \\
MiMo & -0.063 & 0.036 & 0.027 \\
\bottomrule
\end{tabular}
\end{table}

\FloatBarrier
\section{Low-prior subset sensitivity}
\label{app:low_prior_subset}

This appendix repeats the main effort, recovery, ordering-stability, and affinity
diagnostics on the ten low-prior questions: those with question-only
three-trial agreement below 90\% in the released controls. The subset contains
five questions from each task. Accuracy and all-correct gates are recomputed only on these low-prior
questions after the scoring audit. Inspection effort is still recorded at the
whole find-run level, so the effort-based plots assign the full run-level
inspection bytes to the subset.

\begin{figure}[!htbp]
\centering
\includegraphics[width=\textwidth]{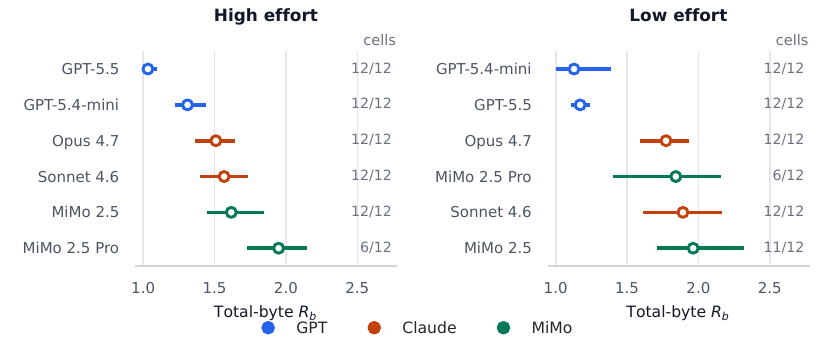}
\caption{Low-prior analogue of Figure~\ref{fig:builderR}. Conditional
inspection effort, $R_b$, is computed on cells where the finder answered all
audited low-prior questions for the task correctly at least once. Lower values
mean less conditional inspection after the subset recovery gate; the cells
column reports contributing finder--task cells out of 12. Bars are computed as in
Figure~\ref{fig:builderR}.}
\label{fig:low_prior_builderR}
\end{figure}

\begin{table}[H]
\centering
\caption{Low-prior analogue of Table~\ref{tab:recovery}. The recovery
gate uses audited low-prior subset accuracy. Entries report task-specific mean
answer recovery, with contributing finder cells shown in parentheses.}
\label{tab:low_prior_recovery}
\begin{minipage}[t]{0.49\textwidth}
\centering
\textbf{High effort}\\[0.25em]
\scriptsize
\setlength{\tabcolsep}{2.7pt}
\begin{tabular}{@{}lcc@{}}
\toprule
Builder & \texttt{minidb} & \texttt{nanoweb} \\
\midrule
GPT-5.5 & 95.0\% (6/6) & 100.0\% (6/6) \\
GPT-5.4-mini & 95.6\% (6/6) & 100.0\% (6/6) \\
Opus 4.7 & 99.4\% (6/6) & 100.0\% (6/6) \\
Sonnet 4.6 & 96.7\% (6/6) & 100.0\% (6/6) \\
MiMo 2.5 Pro & 100.0\% (6/6) & -- (0/6) \\
MiMo 2.5 & 88.7\% (6/6) & 97.8\% (6/6) \\
\bottomrule
\end{tabular}
\end{minipage}\hfill
\begin{minipage}[t]{0.49\textwidth}
\centering
\textbf{Low effort}\\[0.25em]
\scriptsize
\setlength{\tabcolsep}{2.7pt}
\begin{tabular}{@{}lcc@{}}
\toprule
Builder & \texttt{minidb} & \texttt{nanoweb} \\
\midrule
GPT-5.5 & 96.4\% (6/6) & 100.0\% (6/6) \\
GPT-5.4-mini & 98.9\% (6/6) & 98.3\% (6/6) \\
Opus 4.7 & 98.3\% (6/6) & 97.8\% (6/6) \\
Sonnet 4.6 & 98.1\% (6/6) & 98.3\% (6/6) \\
MiMo 2.5 Pro & -- (0/6) & 100.0\% (6/6) \\
MiMo 2.5 & 95.4\% (6/6) & 96.7\% (6/6) \\
\bottomrule
\end{tabular}
\end{minipage}

\end{table}

\begin{figure}[!htbp]
\centering
\includegraphics[width=0.82\textwidth]{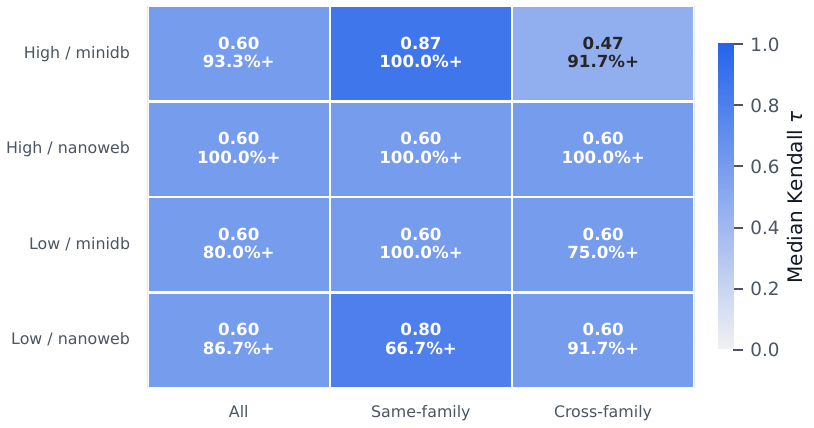}
\caption{Low-prior analogue of Figure~\ref{fig:rankstab}. Pairwise finder
agreement is computed on total-byte orderings in cells where the audited
low-prior subset is recovered perfectly at least once.}
\label{fig:low_prior_rankstab}
\end{figure}

\clearpage
\begin{figure}[H]
\centering
\includegraphics[width=\textwidth]{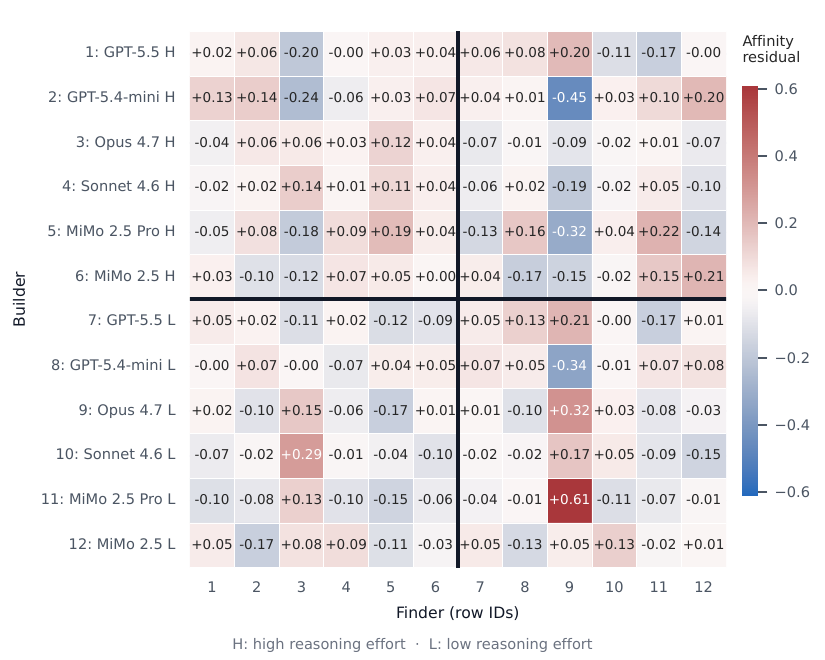}
\caption{Low-prior analogue of Figure~\ref{fig:affinity_residual}. Residuals
use the same matrix normalization as the main affinity diagnostic, with recovery
scores recomputed on the audited low-prior question subset.}
\label{fig:low_prior_affinity_residual}
\end{figure}

\subsection{Option-free check}
\label{app:option_free}

This additional check asks eight of the ten low-prior questions without answer
choices and grades the free-text answers against a rubric. The questions are
\texttt{q\_db\_01}, \texttt{q\_db\_05}, \texttt{q\_db\_06}, \texttt{q\_db\_15},
\texttt{q\_nw\_01}, \texttt{q\_nw\_04}, \texttt{q\_nw\_07}, and
\texttt{q\_nw\_10}. The other two, \texttt{q\_db\_07} and \texttt{q\_nw\_11}, are
negative-contrast items that ask for the inconsistent option and therefore
require visible choices.

\paragraph{Design.}
The grading rubric and a deterministic artifact rule were fixed before any
session ran. The rule takes, per task, the compile-pass artifacts that are
implemented-gold on all 15 questions under the released labels and whose builder
family differs from both finder families, and keeps the lexicographically first
build run identifier. It selected a MiMo 2.5-low artifact for
\texttt{scratch\_minidb} and a MiMo 2.5-high artifact for
\texttt{scratch\_nanoweb}. GPT-5.5-high and Opus 4.7-high answered under two
conditions over three trials. With artifact access, the workspace was a copy of
the selected artifact, and each answer had to cite the source files that
implement the behavior it describes; in the question-only condition, the
workspace held only a marker file, and answers carried no evidence fields. The
24 sessions of four questions give 96 responses and used 9.85M vendor tokens. As
in the main runs, session workspaces were directories inside the repository
checkout, and the runner's copy of the rubric sat two directory levels above
each workspace. The option-free sessions ran in July 2026, and the matched
multiple-choice comparison uses the released find runs of the same finders on
the same artifacts and questions, recorded in May 2026.

\paragraph{Grading.}
The author graded all 96 answers against the rubric in two passes, with no
partial credit, and froze each pass. The grading sheet listed the answers in a
fixed pseudo-random order, sorted by a SHA-256 hash of a fixed salt with the run
and question identifiers, and omitted condition, finder, trial, and run
identifiers. It showed each claim with its rationale. The rubric hides the
evidence citations during pass 1, but the frozen pass-1 sheet includes them;
because only artifact-condition answers carry citations, the condition was often
inferable from a row. Pass 1 graded correctness
(correct, incorrect, or unanswered). Pass 2 graded whether the cited files
contain the code that implements the described behavior and left all 96 pass-1
correctness labels unchanged. The rubric assigns question-only rows the evidence
label not-applicable; the pass-2 sheet records them as citing no evidence, and
the evidence counts below use artifact-condition rows only. An LLM graded the
same rows under the same rubric as a consistency check. Its labels agree with the
author's on correctness for 82 of the 96 rows, with the author stricter in 12 of
the 14 disagreements, and on evidence for all 96; they are used only for these
agreement counts.

\paragraph{Results.}
Table~\ref{tab:option_free} gives the counts and Table~\ref{tab:option_free_questions}
the per-question results. Free-text answers were correct in 38 of 48 responses
with artifact access (79.2\%) and in 13 of 48 without it (27.1\%), while the
matched multiple-choice arms were correct in all 48 answer rows of each condition;
none of the 12 question-only multiple-choice runs is among the control exclusions
of Section~\ref{sec:setup}. The difference has the same sign for both finders and
both tasks. All 38 correct artifact-condition answers, and all 48
artifact-condition answers in total, were graded as citing sufficient evidence.
Four of the eight questions
(\texttt{q\_db\_06}, \texttt{q\_nw\_04}, \texttt{q\_nw\_07}, and
\texttt{q\_nw\_10}) still meet the low-prior criterion after the exclusion in
Section~\ref{sec:setup}; on them, 18 of 24 answers with artifact access and 3 of
24 without it are correct. The 96 responses are repeated observations of eight
questions on one artifact per task.

\paragraph{Implementation labels of the selected pairs.}
The artifact rule used the released labels, under which both artifacts are
implemented-gold on all 15 questions. The source-trace audit of
Appendix~\ref{app:artifact_audit} later flagged 6 of the 15 pairs on the
\texttt{scratch\_minidb} artifact and 10 of the 15 on the
\texttt{scratch\_nanoweb} artifact; under the adjudicated labels, 9 and 5 of
their 15 pairs remain implemented-gold. Three of the eight option-free pairs
change label: \texttt{q\_db\_05} and \texttt{q\_nw\_07} are among the nine
observed errors, and \texttt{q\_db\_01} changes under the adjudicated labels.
Correctness grading applies the rubric's accept and reject criteria as frozen.
Dropping the two observed-error pairs from
both conditions gives 32 of 36 correct answers with artifact access and 9 of 36
without it; dropping all three gives 28 of 30 and 9 of 30.

\begin{table}[H]
\centering
\small
\caption{Option-free check: correct free-text answers with and without artifact
access. Low-prior status uses the question-only controls after the exclusion in
Section~\ref{sec:setup}. The multiple-choice row uses the released find runs of
the same finders on the same artifacts and questions.}
\label{tab:option_free}
\begin{tabular}{lrrr}
\toprule
Slice & Artifact access & Question only & Difference (pp) \\
\midrule
Option-free, all eight questions & 38/48 & 13/48 & $+52.1$ \\
\quad GPT-5.5-high & 19/24 & 5/24 & $+58.3$ \\
\quad Opus 4.7-high & 19/24 & 8/24 & $+45.8$ \\
\quad \texttt{scratch\_minidb} & 18/24 & 5/24 & $+54.2$ \\
\quad \texttt{scratch\_nanoweb} & 20/24 & 8/24 & $+50.0$ \\
\quad Still low-prior (4 questions) & 18/24 & 3/24 & $+62.5$ \\
\quad Without 2 observed-error pairs & 32/36 & 9/36 & $+63.9$ \\
\quad Without 3 pairs relabeled after adjudication & 28/30 & 9/30 & $+63.3$ \\
\midrule
Multiple choice, same items & 48/48 & 48/48 & -- \\
\bottomrule
\end{tabular}
\end{table}

\begin{table}[!htbp]
\centering
\small
\caption{Option-free correctness per question; each cell has six responses (two
finders, three trials). All eight pairs are implemented-gold under the released
labels.}
\label{tab:option_free_questions}
\begin{tabular}{lrrcl}
\toprule
Question & Artifact access & Question only & Still low-prior & Adjudicated label \\
\midrule
\texttt{q\_db\_01} & 4/6 & 0/6 & no & absent/ambiguous \\
\texttt{q\_db\_05} & 4/6 & 1/6 & no & implemented-non-gold \\
\texttt{q\_db\_06} & 4/6 & 0/6 & yes & implemented-gold \\
\texttt{q\_db\_15} & 6/6 & 4/6 & no & implemented-gold \\
\texttt{q\_nw\_01} & 6/6 & 5/6 & no & implemented-gold \\
\texttt{q\_nw\_04} & 6/6 & 0/6 & yes & implemented-gold \\
\texttt{q\_nw\_07} & 2/6 & 3/6 & yes & absent/ambiguous \\
\texttt{q\_nw\_10} & 6/6 & 0/6 & yes & implemented-gold \\
\bottomrule
\end{tabular}
\end{table}

\section{Resource usage diagnostics}
\label{app:usage}

Vendor token counts are reported as diagnostics. They use observed vendor-token totals from the efficiency records, as exported
to the canonical run tables, including subagent calls. Table~\ref{tab:token_usage}
reports the input/output split for the artifact-present formal panel, including
compile-failed artifacts and the artifact-conditioned find runs they triggered,
because compute diagnostics account for spent model calls across the full
artifact-present release. Input and output are split so their sum equals
the token total used elsewhere in this appendix; input includes cache-read and
cache-creation categories when the provider reports them separately, and output
includes reasoning tokens when reported. The 144 control find runs add 19.3M
tokens in total (9.9M spec-only and 9.4M question-only). The conditional
builder-effort diagnostic remains within-finder relative inspection effort over
novel inspection bytes.
All reported runs were orchestrated on a MacBookPro18,3 with an Apple M1 Pro
CPU, 10 cores, 16 GB RAM, macOS 15.6, and arm64 architecture. Local compute was
used for orchestration, Cargo compile probes, and deterministic analysis; model
inference ran on provider infrastructure. Run records include
\texttt{wall\_clock\_seconds}; token totals are reported as the stable compute
diagnostic because live rerun latency depends on provider scheduling and model
serving.

\begin{table}[!htbp]
\centering
\scriptsize
\caption{Vendor token usage diagnostic by role and agent configuration in the
artifact-present formal panel, including subagent calls. Token columns are totals in millions except
mean/run, which is thousands of total tokens per run.}
\label{tab:token_usage}
\begin{tabular}{llrrrrr}
\toprule
Role & Agent & Runs & Input & Output & Total & Mean/run \\
\midrule
Build & Claude Opus 4.7-high & 4 & 25.8 & 0.46 & 26.3 & 6566 \\
Build & Claude Opus 4.7-low & 4 & 19.0 & 0.32 & 19.3 & 4821 \\
Build & Claude Sonnet 4.6-high & 4 & 31.1 & 0.41 & 31.5 & 7867 \\
Build & Claude Sonnet 4.6-low & 4 & 20.1 & 0.29 & 20.4 & 5093 \\
Build & GPT-5.5-high & 4 & 12.7 & 0.16 & 12.9 & 3217 \\
Build & GPT-5.5-low & 4 & 7.6 & 0.10 & 7.7 & 1922 \\
Build & GPT-5.4 Mini-high & 4 & 29.5 & 0.36 & 29.9 & 7471 \\
Build & GPT-5.4 Mini-low & 4 & 8.8 & 0.09 & 8.9 & 2231 \\
Build & MiMo 2.5 Pro-high & 4 & 57.9 & 0.40 & 58.3 & 14577 \\
Build & MiMo 2.5 Pro-low & 4 & 38.6 & 0.35 & 38.9 & 9731 \\
Build & MiMo 2.5-high & 4 & 61.3 & 0.36 & 61.7 & 15425 \\
Build & MiMo 2.5-low & 4 & 42.3 & 0.26 & 42.6 & 10652 \\
\midrule
Find & Claude Opus 4.7-high & 144 & 134.7 & 1.12 & 135.9 & 943 \\
Find & Claude Opus 4.7-low & 144 & 76.1 & 0.51 & 76.6 & 532 \\
Find & Claude Sonnet 4.6-high & 144 & 103.8 & 1.98 & 105.8 & 734 \\
Find & Claude Sonnet 4.6-low & 144 & 97.6 & 1.51 & 99.1 & 688 \\
Find & GPT-5.5-high & 144 & 79.5 & 0.96 & 80.5 & 559 \\
Find & GPT-5.5-low & 144 & 55.5 & 0.62 & 56.1 & 389 \\
Find & GPT-5.4 Mini-high & 144 & 93.9 & 1.67 & 95.6 & 664 \\
Find & GPT-5.4 Mini-low & 144 & 48.0 & 0.65 & 48.6 & 338 \\
Find & MiMo 2.5 Pro-high & 144 & 58.5 & 1.08 & 59.6 & 414 \\
Find & MiMo 2.5 Pro-low & 144 & 59.5 & 1.06 & 60.6 & 421 \\
Find & MiMo 2.5-high & 144 & 56.8 & 0.98 & 57.8 & 401 \\
Find & MiMo 2.5-low & 144 & 56.5 & 0.93 & 57.4 & 399 \\
\bottomrule
\end{tabular}
\end{table}

\section{Benchmark card}
\label{app:benchcard}

\paragraph{Intended use.}
The benchmark is intended for comparing generated artifacts within a calibrated
finder-task panel by the recoverability of specification-traced design intent,
selecting and validating finder panels, and auditing generated repositories for
agent-facing legibility. Private task packs are intended for
contamination-resistant evaluation under the same protocol.

\paragraph{Out-of-scope use.}
Out-of-scope uses include certifying deployment safety, security, runtime
correctness beyond the recorded build-validation fields, human maintainability,
public-task-pack leaderboards, and single-axis universal rankings of coding
models.

\paragraph{Data composition.}
The released task pack contains \texttt{scratch\_minidb} and
\texttt{scratch\_nanoweb}, each with a hidden specification, canonical
intended behaviors and design choices, and 15 specification-traced MCQs. The
canonical artifact-present release contains 48 build rows and 1728
artifact-conditioned find rows. The reported compile-pass primary panel
contains 41 build rows and 1476 artifact-conditioned find rows; the released
tables also contain 144 control find rows. After the scoring audit, the primary
scored recovery set contains 21312 finder-answer rows. Seven compile-failed artifacts
remain in the release as validity-stress diagnostics. The release
includes derived analysis reports.

\paragraph{Quality control.}
Specifications are schema-validated and manually reviewed by the author.
Question banks must reference canonical intended behaviors and design choices
and pass duplicate-option checks; the author also inspects the questions,
distractors, trace annotations, and gold answers for ambiguity and traceability.
The question banks were written with LLM assistance and reviewed by the author.
Generated artifacts receive build-validation checks during the build phase; the
reported primary analysis gates on compile-pass artifacts, audits
artifact-question implementation status, and reports excluded artifacts
separately.
Canonical tables are produced from append-only run records by deterministic
transforms.

\paragraph{Metrics.}
Metrics are exact-match recovery accuracy as a validity check, within-finder
relative inspection effort $R_b$ over novel inspection bytes on all-correct
cells, and a co-reported task-level recovery gate that penalizes partial
recovery without mixing in effort. Diagnostics include spec-only and question-only controls, prior-lift
views, compile-failed stress diagnostics, rank stability, builder--finder
affinity views, artifact-evidence audits, low-prior sensitivity views, and
token-usage summaries. In the current high-prior task
pack, accuracy is a gate and calibration signal, and conditional effort carries
the main discriminative signal. Conditional inspection effort is interpreted as a proxy
for agent-facing artifact legibility only after recovery and stability gates are
satisfied. Affinity is panel-local and diagnostic.

\paragraph{Known biases.}
The current task pack is single-language and covers two task families, while the
generated artifacts remain substantial: among source-bearing final build cells,
the median artifact contains roughly 4.2k source LOC, and many span 2k--9k LOC.
This scale makes repeated builder--finder trials costly; the canonical
artifact-present release consumed more than 1.1B vendor-reported tokens across
formal and control runs, while the compile-pass primary panel uses roughly
1.095B. Questions
intentionally preserve ordinary software-engineering priors, and controls show
that those priors are strong in this release: question-only accuracy is 97.2\%. The intended interpretation is
panel-conditioned and prior-conditioned, including the finder policies and CLI
harnesses used here. The public task pack has panel-local scope; private task
packs are the intended route for contamination resistance and broader coverage.

\paragraph{Maintenance.}
\label{sec:maintenance}
Task packs and tables are versioned. New task packs can be added without
changing the record schema. Corrections are recorded as new evidence layers
with prior records preserved.

\paragraph{Pack admission.}
New task packs may use LLMs to draft specifications, questions, and distractors,
under admission rules fixed before any candidate builder is evaluated. A drafted
item enters a pack only after question-only and option-free prior screening and
human review of its gold answer. In new packs, the specification gold is
counterbalanced across equally plausible design variants of a task family, so
that the gold does not systematically coincide with the most common engineering
design. New packs retain their drafting and distractor-selection histories;
these histories were not retained for the current pack. Each pack needs its own
prior and finder calibration. Specifications
stay compact, because longer specifications shift the measurement toward builder
instruction following and away from the artifact-side communication the protocol
targets. Scale is constrained mainly by evaluation cost: every admitted task is
built by every builder configuration over repeated trials and inspected by a
replicated finder panel.

\newpage
\section*{NeurIPS Paper Checklist}

\begin{enumerate}

\Needspace{6\baselineskip}
\item {\bf Claims}\nopagebreak[4]
    \item[] Question: Do the main claims made in the abstract and introduction accurately reflect the paper's contributions and scope?
    \item[] Answer: \answerYes{}
    \item[] Justification: The abstract and introduction state the protocol, construct, reported metrics, controls, and release. The main text scopes interpretation to the released high-prior panel and treats effort and affinity as panel-local diagnostics.
    \item[] Guidelines:
    \begin{itemize}
        \item The answer \answerNA{} means that the abstract and introduction do not include the claims made in the paper.
        \item The abstract and/or introduction should clearly state the claims made, including the contributions made in the paper and important assumptions and limitations. A \answerNo{} or \answerNA{} answer to this question will not be perceived well by the reviewers.
        \item The claims made should match theoretical and experimental results, and reflect how much the results can be expected to generalize to other settings.
        \item It is fine to include aspirational goals as motivation as long as it is clear that these goals are not attained by the paper.
    \end{itemize}

\Needspace{6\baselineskip}
\item {\bf Limitations}\nopagebreak[4]
    \item[] Question: Does the paper discuss the limitations of the work performed by the authors?
    \item[] Answer: \answerYes{}
    \item[] Justification: Sections~\ref{sec:controls} and~\ref{sec:validity-scope} discuss strong priors, finder policies, panel and task scope, contamination, and the limits of agent-facing recoverability; Section~\ref{sec:release} and Appendix~\ref{app:reproducibility} discuss model drift. The paper also separates compile/layout validation from full behavioral conformance.
    \item[] Guidelines:
    \begin{itemize}
        \item The answer \answerNA{} means that the paper has no limitation while the answer \answerNo{} means that the paper has limitations, but those are not discussed in the paper.
        \item The authors are encouraged to create a separate ``Limitations'' section in their paper.
        \item The paper should point out any strong assumptions and how robust the results are to violations of these assumptions (e.g., independence assumptions, noiseless settings, model well-specification, asymptotic approximations only holding locally). The authors should reflect on how these assumptions might be violated in practice and what the implications would be.
        \item The authors should reflect on the scope of the claims made, e.g., if the approach was only tested on a few datasets or with a few runs. In general, empirical results often depend on implicit assumptions, which should be articulated.
        \item The authors should reflect on the factors that influence the performance of the approach. For example, a facial recognition algorithm may perform poorly when image resolution is low or images are taken in low lighting. Or a speech-to-text system might not be used reliably to provide closed captions for online lectures because it fails to handle technical jargon.
        \item The authors should discuss the computational efficiency of the proposed algorithms and how they scale with dataset size.
        \item If applicable, the authors should discuss possible limitations of their approach to address problems of privacy and fairness.
        \item While the authors might fear that complete honesty about limitations might be used by reviewers as grounds for rejection, a worse outcome might be that reviewers discover limitations that aren't acknowledged in the paper. The authors should use their best judgment and recognize that individual actions in favor of transparency play an important role in developing norms that preserve the integrity of the community. Reviewers will be specifically instructed to not penalize honesty concerning limitations.
    \end{itemize}

\Needspace{6\baselineskip}
\item {\bf Theory assumptions and proofs}\nopagebreak[4]
    \item[] Question: For each theoretical result, does the paper provide the full set of assumptions and a complete (and correct) proof?
    \item[] Answer: \answerNA{}
    \item[] Justification: The paper contains no formal theoretical results; $R_b$ is an operational metric defined in Section~\ref{sec:metrics}.
    \item[] Guidelines:
    \begin{itemize}
        \item The answer \answerNA{} means that the paper does not include theoretical results.
        \item All the theorems, formulas, and proofs in the paper should be numbered and cross-referenced.
        \item All assumptions should be clearly stated or referenced in the statement of any theorems.
        \item The proofs can either appear in the main paper or the supplemental material, but if they appear in the supplemental material, the authors are encouraged to provide a short proof sketch to provide intuition.
        \item Inversely, any informal proof provided in the core of the paper should be complemented by formal proofs provided in appendix or supplemental material.
        \item Theorems and Lemmas that the proof relies upon should be properly referenced.
    \end{itemize}

    \Needspace{6\baselineskip}
\item {\bf Experimental result reproducibility}\nopagebreak[4]
    \item[] Question: Does the paper fully disclose all the information needed to reproduce the main experimental results of the paper to the extent that it affects the main claims and/or conclusions of the paper (regardless of whether the code and data are provided or not)?
    \item[] Answer: \answerYes{}
    \item[] Justification: Sections~\ref{sec:empirical},~\ref{app:impl}, and~\ref{app:reproducibility} report tasks, agents, trials, seeds, hashes, manifests, and toolchain versions. Stored records and deterministic transforms reproduce the analysis; live reruns are subject to provider and model drift.
    \item[] Guidelines:
    \begin{itemize}
        \item The answer \answerNA{} means that the paper does not include experiments.
        \item If the paper includes experiments, a \answerNo{} answer to this question will not be perceived well by the reviewers: Making the paper reproducible is important, regardless of whether the code and data are provided or not.
        \item If the contribution is a dataset and\slash or model, the authors should describe the steps taken to make their results reproducible or verifiable.
        \item Depending on the contribution, reproducibility can be accomplished in various ways. For example, if the contribution is a novel architecture, describing the architecture fully might suffice, or if the contribution is a specific model and empirical evaluation, it may be necessary to either make it possible for others to replicate the model with the same dataset, or provide access to the model. In general. releasing code and data is often one good way to accomplish this, but reproducibility can also be provided via detailed instructions for how to replicate the results, access to a hosted model (e.g., in the case of a large language model), releasing of a model checkpoint, or other means that are appropriate to the research performed.
        \item While NeurIPS does not require releasing code, the conference does require all submissions to provide some reasonable avenue for reproducibility, which may depend on the nature of the contribution. For example
        \begin{enumerate}
            \item If the contribution is primarily a new algorithm, the paper should make it clear how to reproduce that algorithm.
            \item If the contribution is primarily a new model architecture, the paper should describe the architecture clearly and fully.
            \item If the contribution is a new model (e.g., a large language model), then there should either be a way to access this model for reproducing the results or a way to reproduce the model (e.g., with an open-source dataset or instructions for how to construct the dataset).
            \item We recognize that reproducibility may be tricky in some cases, in which case authors are welcome to describe the particular way they provide for reproducibility. In the case of closed-source models, it may be that access to the model is limited in some way (e.g., to registered users), but it should be possible for other researchers to have some path to reproducing or verifying the results.
        \end{enumerate}
    \end{itemize}

\Needspace{6\baselineskip}
\item {\bf Open access to data and code}\nopagebreak[4]
    \item[] Question: Does the paper provide open access to the data and code, with sufficient instructions to faithfully reproduce the main experimental results, as described in supplemental material?
    \item[] Answer: \answerYes{}
    \item[] Justification: Section~\ref{sec:release} states that the harness, tasks, question banks, artifacts, records, tables, reports, scripts, metadata, licenses, and evidence audit are released at the code and dataset URLs given there. The appendices document manifests, configurations, and deterministic analysis.
    \item[] Guidelines:
    \begin{itemize}
        \item The answer \answerNA{} means that paper does not include experiments requiring code.
        \item Please see the NeurIPS code and data submission guidelines (\url{https://neurips.cc/public/guides/CodeSubmissionPolicy}) for more details.
        \item While we encourage the release of code and data, we understand that this might not be possible, so \answerNo{} is an acceptable answer. Papers cannot be rejected simply for not including code, unless this is central to the contribution (e.g., for a new open-source benchmark).
        \item The instructions should contain the exact command and environment needed to run to reproduce the results. See the NeurIPS code and data submission guidelines (\url{https://neurips.cc/public/guides/CodeSubmissionPolicy}) for more details.
        \item The authors should provide instructions on data access and preparation, including how to access the raw data, preprocessed data, intermediate data, and generated data, etc.
        \item The authors should provide scripts to reproduce all experimental results for the new proposed method and baselines. If only a subset of experiments are reproducible, they should state which ones are omitted from the script and why.
        \item At submission time, to preserve anonymity, the authors should release anonymized versions (if applicable).
        \item Providing as much information as possible in supplemental material (appended to the paper) is recommended, but including URLs to data and code is permitted.
    \end{itemize}

\Needspace{6\baselineskip}
\item {\bf Experimental setting/details}\nopagebreak[4]
    \item[] Question: Does the paper specify all the training and test details (e.g., data splits, hyperparameters, how they were chosen, type of optimizer) necessary to understand the results?
    \item[] Answer: \answerYes{}
    \item[] Justification: Section~\ref{sec:empirical} describes the two tasks, agent panel, build/find trial counts, scoring audit, and effort metrics. Tables~\ref{tab:agents} and~\ref{tab:execution_stack} give agent and toolchain configuration.
    \item[] Guidelines:
    \begin{itemize}
        \item The answer \answerNA{} means that the paper does not include experiments.
        \item The experimental setting should be presented in the core of the paper to a level of detail that is necessary to appreciate the results and make sense of them.
        \item The full details can be provided either with the code, in appendix, or as supplemental material.
    \end{itemize}

\Needspace{12\baselineskip}
\item {\bf Experiment statistical significance}\nopagebreak[4]
    \item[] Question: Does the paper report error bars suitably and correctly defined or other appropriate information about the statistical significance of the experiments?
    \item[] Answer: \answerYes{}
    \item[] Justification: Figures~\ref{fig:builderR} and~\ref{fig:low_prior_builderR} show 95\% percentile bootstrap intervals for the builder effort score $R_b$, from 2{,}000 resamples of the 12 finder--task cells of each effort panel. Tables~\ref{tab:prior-stress}, \ref{tab:low-prior-evidence-summary}, and~\ref{tab:prior-stress-by-task} report 95\% percentile bootstrap intervals for accuracy differences from the question-only control, from 2{,}000 resamples of task--question clusters over matched finder, task, and trial strata. Appendix~\ref{app:metric_details} states both procedures. The intervals capture variability across finder--task cells or questions for the fixed generated artifacts and finder panel.
    \item[] Guidelines:
    \begin{itemize}
        \item The answer \answerNA{} means that the paper does not include experiments.
        \item The authors should answer \answerYes{} if the results are accompanied by error bars, confidence intervals, or statistical significance tests, at least for the experiments that support the main claims of the paper.
        \item The factors of variability that the error bars are capturing should be clearly stated (for example, train/test split, initialization, random drawing of some parameter, or overall run with given experimental conditions).
        \item The method for calculating the error bars should be explained (closed form formula, call to a library function, bootstrap, etc.)
        \item The assumptions made should be given (e.g., Normally distributed errors).
        \item It should be clear whether the error bar is the standard deviation or the standard error of the mean.
        \item It is OK to report 1-sigma error bars, but one should state it. The authors should preferably report a 2-sigma error bar than state that they have a 96\% CI, if the hypothesis of Normality of errors is not verified.
        \item For asymmetric distributions, the authors should be careful not to show in tables or figures symmetric error bars that would yield results that are out of range (e.g., negative error rates).
        \item If error bars are reported in tables or plots, the authors should explain in the text how they were calculated and reference the corresponding figures or tables in the text.
    \end{itemize}

\Needspace{6\baselineskip}
\item {\bf Experiments compute resources}\nopagebreak[4]
    \item[] Question: For each experiment, does the paper provide sufficient information on the computer resources (type of compute workers, memory, time of execution) needed to reproduce the experiments?
    \item[] Answer: \answerYes{}
    \item[] Justification: Section~\ref{sec:empirical} reports run counts. Appendix~\ref{app:usage} reports vendor token usage, local MacBook hardware, CPU-only local work, provider-side inference, and recorded wall-clock fields.
    \item[] Guidelines:
    \begin{itemize}
        \item The answer \answerNA{} means that the paper does not include experiments.
        \item The paper should indicate the type of compute workers CPU or GPU, internal cluster, or cloud provider, including relevant memory and storage.
        \item The paper should provide the amount of compute required for each of the individual experimental runs as well as estimate the total compute.
        \item The paper should disclose whether the full research project required more compute than the experiments reported in the paper (e.g., preliminary or failed experiments that didn't make it into the paper).
    \end{itemize}

\Needspace{6\baselineskip}
\item {\bf Code of ethics}\nopagebreak[4]
    \item[] Question: Does the research conducted in the paper conform, in every respect, with the NeurIPS Code of Ethics \url{https://neurips.cc/public/EthicsGuidelines}?
    \item[] Answer: \answerYes{}
    \item[] Justification: Section~\ref{sec:release} states that the tasks are synthetic and that no human-subject data are used, and recommends OS-level sandboxing with resource limits for coding-agent runs.
    \item[] Guidelines:
    \begin{itemize}
        \item The answer \answerNA{} means that the authors have not reviewed the NeurIPS Code of Ethics.
        \item If the authors answer \answerNo, they should explain the special circumstances that require a deviation from the Code of Ethics.
        \item The authors should make sure to preserve anonymity (e.g., if there is a special consideration due to laws or regulations in their jurisdiction).
    \end{itemize}

\Needspace{6\baselineskip}
\item {\bf Broader impacts}\nopagebreak[4]
    \item[] Question: Does the paper discuss both potential positive societal impacts and negative societal impacts of the work performed?
    \item[] Answer: \answerYes{}
    \item[] Justification: Section~\ref{sec:release} and Appendix~\ref{app:benchcard} describe intended diagnostic uses, sandboxing recommendations, contamination concerns, and out-of-scope uses such as deployment-safety certification and universal model rankings.
    \item[] Guidelines:
    \begin{itemize}
        \item The answer \answerNA{} means that there is no societal impact of the work performed.
        \item If the authors answer \answerNA{} or \answerNo, they should explain why their work has no societal impact or why the paper does not address societal impact.
        \item Examples of negative societal impacts include potential malicious or unintended uses (e.g., disinformation, generating fake profiles, surveillance), fairness considerations (e.g., deployment of technologies that could make decisions that unfairly impact specific groups), privacy considerations, and security considerations.
        \item The conference expects that many papers will be foundational research and not tied to particular applications, let alone deployments. However, if there is a direct path to any negative applications, the authors should point it out. For example, it is legitimate to point out that an improvement in the quality of generative models could be used to generate Deepfakes for disinformation. On the other hand, it is not needed to point out that a generic algorithm for optimizing neural networks could enable people to train models that generate Deepfakes faster.
        \item The authors should consider possible harms that could arise when the technology is being used as intended and functioning correctly, harms that could arise when the technology is being used as intended but gives incorrect results, and harms following from (intentional or unintentional) misuse of the technology.
        \item If there are negative societal impacts, the authors could also discuss possible mitigation strategies (e.g., gated release of models, providing defenses in addition to attacks, mechanisms for monitoring misuse, mechanisms to monitor how a system learns from feedback over time, improving the efficiency and accessibility of ML).
    \end{itemize}

\Needspace{6\baselineskip}
\item {\bf Safeguards}\nopagebreak[4]
    \item[] Question: Does the paper describe safeguards that have been put in place for responsible release of data or models that have a high risk for misuse (e.g., pre-trained language models, image generators, or scraped datasets)?
    \item[] Answer: \answerNA{}
    \item[] Justification: The release contains benchmark artifacts and harness code, not a generative model or scraped dataset of the kind targeted by this question.
    \item[] Guidelines:
    \begin{itemize}
        \item The answer \answerNA{} means that the paper poses no such risks.
        \item Released models that have a high risk for misuse or dual-use should be released with necessary safeguards to allow for controlled use of the model, for example by requiring that users adhere to usage guidelines or restrictions to access the model or implementing safety filters.
        \item Datasets that have been scraped from the Internet could pose safety risks. The authors should describe how they avoided releasing unsafe images.
        \item We recognize that providing effective safeguards is challenging, and many papers do not require this, but we encourage authors to take this into account and make a best faith effort.
    \end{itemize}

\Needspace{6\baselineskip}
\item {\bf Licenses for existing assets}\nopagebreak[4]
    \item[] Question: Are the creators or original owners of assets (e.g., code, data, models), used in the paper, properly credited and are the license and terms of use explicitly mentioned and properly respected?
    \item[] Answer: \answerYes{}
    \item[] Justification: Prior benchmarks are cited for context and are not redistributed. The paper identifies the model APIs and CLIs used, states that they are accessed under provider terms, and releases no model weights or provider code.
    \item[] Guidelines:
    \begin{itemize}
        \item The answer \answerNA{} means that the paper does not use existing assets.
        \item The authors should cite the original paper that produced the code package or dataset.
        \item The authors should state which version of the asset is used and, if possible, include a URL.
        \item The name of the license (e.g., CC-BY 4.0) should be included for each asset.
        \item For scraped data from a particular source (e.g., website), the copyright and terms of service of that source should be provided.
        \item If assets are released, the license, copyright information, and terms of use in the package should be provided. For popular datasets, \url{paperswithcode.com/datasets} has curated licenses for some datasets. Their licensing guide can help determine the license of a dataset.
        \item For existing datasets that are re-packaged, both the original license and the license of the derived asset (if it has changed) should be provided.
        \item If this information is not available online, the authors are encouraged to reach out to the asset's creators.
    \end{itemize}

\Needspace{6\baselineskip}
\item {\bf New assets}\nopagebreak[4]
    \item[] Question: Are new assets introduced in the paper well documented and is the documentation provided alongside the assets?
    \item[] Answer: \answerYes{}
    \item[] Justification: The released harness, tasks, question banks, generated artifacts, records, tables, reports, metadata, and audits are new assets. They are documented in Appendices~\ref{app:impl},~\ref{app:reproducibility}, and~\ref{app:benchcard}, with licenses included in the release (code: Apache License 2.0; data: CC BY 4.0).
    \item[] Guidelines:
    \begin{itemize}
        \item The answer \answerNA{} means that the paper does not release new assets.
        \item Researchers should communicate the details of the dataset\slash code\slash model as part of their submissions via structured templates. This includes details about training, license, limitations, etc.
        \item The paper should discuss whether and how consent was obtained from people whose asset is used.
        \item At submission time, remember to anonymize your assets (if applicable). You can either create an anonymized URL or include an anonymized zip file.
    \end{itemize}

\Needspace{6\baselineskip}
\item {\bf Crowdsourcing and research with human subjects}\nopagebreak[4]
    \item[] Question: For crowdsourcing experiments and research with human subjects, does the paper include the full text of instructions given to participants and screenshots, if applicable, as well as details about compensation (if any)?
    \item[] Answer: \answerNA{}
    \item[] Justification: The benchmark uses synthetic software tasks and generated code artifacts; no crowdsourcing or human-subject study is involved.
    \item[] Guidelines:
    \begin{itemize}
        \item The answer \answerNA{} means that the paper does not involve crowdsourcing nor research with human subjects.
        \item Including this information in the supplemental material is fine, but if the main contribution of the paper involves human subjects, then as much detail as possible should be included in the main paper.
        \item According to the NeurIPS Code of Ethics, workers involved in data collection, curation, or other labor should be paid at least the minimum wage in the country of the data collector.
    \end{itemize}

\Needspace{6\baselineskip}
\item {\bf Institutional review board (IRB) approvals or equivalent for research with human subjects}\nopagebreak[4]
    \item[] Question: Does the paper describe potential risks incurred by study participants, whether such risks were disclosed to the subjects, and whether Institutional Review Board (IRB) approvals (or an equivalent approval/review based on the requirements of your country or institution) were obtained?
    \item[] Answer: \answerNA{}
    \item[] Justification: No human-subject research is conducted; the study evaluates coding-agent software artifacts.
    \item[] Guidelines:
    \begin{itemize}
        \item The answer \answerNA{} means that the paper does not involve crowdsourcing nor research with human subjects.
        \item Depending on the country in which research is conducted, IRB approval (or equivalent) may be required for any human subjects research. If you obtained IRB approval, you should clearly state this in the paper.
        \item We recognize that the procedures for this may vary significantly between institutions and locations, and we expect authors to adhere to the NeurIPS Code of Ethics and the guidelines for their institution.
        \item For initial submissions, do not include any information that would break anonymity (if applicable), such as the institution conducting the review.
    \end{itemize}

\Needspace{6\baselineskip}
\item {\bf Declaration of LLM usage}\nopagebreak[4]
    \item[] Question: Does the paper describe the usage of LLMs if it is an important, original, or non-standard component of the core methods in this research? Note that if the LLM is used only for writing, editing, or formatting purposes and does \emph{not} impact the core methodology, scientific rigor, or originality of the research, declaration is not required.
    \item[] Answer: \answerYes{}
    \item[] Justification: LLM-based coding agents are the builders and finders studied by the protocol; Sections~\ref{sec:empirical} and~\ref{app:reproducibility} report the panel and configurations. The question banks were written with LLM assistance and reviewed by the author (Appendix~\ref{app:benchcard}). LLMs were also used in the implementation-label audit, for a model-assisted source trace of the scored artifact-question pairs and for drafting decisions on flagged pairs, which the author reviewed and approved (Appendix~\ref{app:artifact_audit}). In the option-free check, an LLM also graded the free-text answers for a consistency check; its labels enter only an agreement count (Appendix~\ref{app:option_free}).
    \item[] Guidelines:
    \begin{itemize}
        \item The answer \answerNA{} means that the core method development in this research does not involve LLMs as any important, original, or non-standard components.
        \item Please refer to our LLM policy in the NeurIPS handbook for what should or should not be described.
    \end{itemize}

\end{enumerate}

\end{document}